\documentclass[twocolumn,showpacs,superscriptaddress,preprintnumbers,floatfix,amsmath,amssymb,prx]{revtex4-1}

\usepackage{graphicx,tabularx,color,epsfig,rotate,ulem,float}
\usepackage{mathbbol}
\usepackage{amsmath}
\usepackage{amssymb}
\usepackage{hyperref}
\usepackage{multirow}
\usepackage{subfigure}
\usepackage{epstopdf}
\usepackage[below]{placeins}
\usepackage{graphicx}
\usepackage{tabularx}

\begin{document}


\title{Exotic magnetic field-induced spin-superstructures in \\ a mixed honeycomb triangular lattice system}

\author{V. Ovidiu Garlea}
 \email{garleao@ornl.gov}
\affiliation{Neutron Scattering Division, Oak Ridge National Laboratory, Oak Ridge, Tennessee 37831, USA}
\author{Liurukara D. Sanjeewa}
\affiliation{Department of Chemistry and Center for Optical Materials Science and Engineering Technologies (COMSET), Clemson University, Clemson, South Carolina 29634-0973, USA}
\affiliation{Materials Science and Technology Division, Oak Ridge National Laboratory, Oak Ridge, Tennessee 37831, USA}
\author{Michael A. McGuire}
\affiliation{Materials Science and Technology Division, Oak Ridge National Laboratory, Oak Ridge, Tennessee 37831, USA}
\author{Cristian D. Batista}
\affiliation{Shull-Wollan Center, Oak Ridge National Laboratory, Oak Ridge, Tennessee 37831, USA}
\affiliation{Department of Physics and Astronomy, University of Tennessee, Knoxville, Tennessee 37996, USA}
\author{Anjana M. Samarakoon}
\affiliation{Shull-Wollan Center, Oak Ridge National Laboratory, Oak Ridge, Tennessee 37831, USA}
\author{David Graf}
\affiliation{National High Magnetic Field Laboratory, Florida State University, Tallahassee, Florida 32310, USA}
\author{Barry Winn}
\author{Feng Ye}
\author{Christina Hoffmann}
\affiliation{Neutron Scattering Division, Oak Ridge National Laboratory, Oak Ridge, Tennessee 37831, USA}
\author{Joseph W. Kolis}
\affiliation{Department of Chemistry and Center for Optical Materials Science and Engineering Technologies (COMSET), Clemson University, Clemson, South Carolina 29634-0973, USA}

\date{\today}

\begin{abstract}
The temperature-magnetic-field phase diagram of the mixed honeycomb triangular lattice system K$_{2}$Mn$_{3}$(VO$_{4}$)$_{2}$CO$_{3}$ is investigated by means of magnetization, heat capacity and neutron scattering measurements. The results indicate that triangular and honeycomb magnetic layers undergo sequential magnetic orderings and act as nearly independent magnetic sublattices. The honeycomb sublattice orders at about 85 K in a Ne\'{e}l-type antiferromagnetic structure, while the triangular sublattice displays two consecutive ordered states at much lower temperatures, 3 K and 2.2 K. The ground state of the triangular sublattice consists of a planar ``Y'' magnetic structure that emerges from an intermediate collinear ``\textit{up-up-down}'' state. Applied magnetic fields parallel or perpendicular to the $c$-axis induce exotic ordered phases characterized by various spin-stacking sequences of triangular layers that yield bilayer, three-layer or four-layer magnetic superstructures. The observed superstructures cannot be explained in the framework of quasi-classical theory based only on nearest-neighbor interlayer coupling and point towards the presence of effective second-nearest-neighbor interactions mediated by fluctuations of the magnetic moments in the honeycomb sublattice.
\end{abstract}


\maketitle

\section{Introduction}

\begin{figure}[tbp]
\includegraphics[width=3.0in]{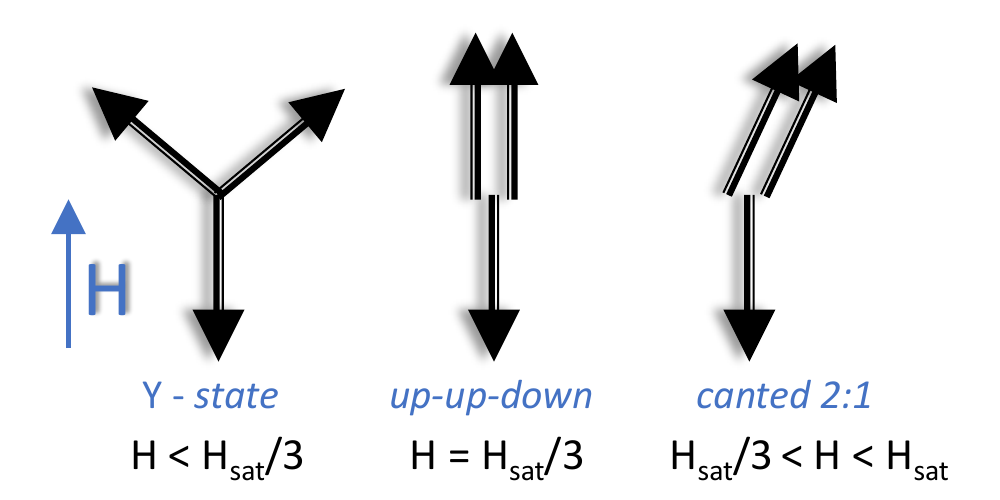}
\caption{\label{spinsconfigs} Planar spin configurations of single-layer triangular lattice antiferromagnet at 0 K and finite magnetic fields.\cite{chubukov}}
\end{figure}

Field-induced magnetic states that occur in layered triangular lattice antiferromagnets (TLA) have been extensively studied and discussed in the context of broken discrete symmetries of the lattice and spin-rotation in the plane perpendicular to the applied field. It is widely recognized that thermal and quantum fluctuations lift the degeneracy of the classical spin configurations of the triangular antiferromagnetic Heisenberg model in a magnetic field. This well-known realization of  \textit{order-by-disorder}\cite{henley} leads to specific planar states: a ``Y''-state or 120$^\circ$ spin configuration with two spins canting ``\textit{up}'' and one pinned in a ``\textit{down}'' direction, a collinear ``\textit{up-up-down}'' ($uud$) state, and a canted ``2:1'' phase that is an oblique version of the $uud$ state.\cite{kawamura,saturation,chubukov,gekht,seabra,seabra1,starykh,interlayer} These planar spin configurations are sketched in Fig.~\ref{spinsconfigs}. An easy-axis anisotropy can also remove the degeneracy and stabilize the same coplanar arrangements. The collinear $uud$ state, which gives a one-third of the saturation magnetization plateau at intermediate fields ($M$=$M_s$/3), breaks the discrete $\mathbb{Z}_3$ symmetry of the lattice. The canted planar ``Y'' and ``2:1'' states that involve $S_x-–S_y$ spin components also break the continuous $U$(1) symmetry of spin rotations about the field axis. The latter spin states with broken mixed symmetries $\mathbb{Z}_3 \otimes U$(1), can be viewed as magnetic supersolid phases that combine superfluid properties with long-range periodicity of solids, as proposed by Liu and Fisher.~\cite{seabra1,supersolid}

Most of existing theoretical studies on TLA assume negligible interplanar interactions. However, the presence of such coupling, as expected in real materials, can cause additional exotic field-induced phase transitions where the relative spin orientations change between adjacent planes to produce interesting magnetic superstructures. The spin-stacking pattern is expected to dramatically depend on the manner of stacking of neighboring layers which can be either eclipsed (adjacent layers related by a simple translation perpendicular to the layer plane) or staggered (shifted by certain in-plane lattice translation) near-neighbor planes, with the latter also enabling frustrated out-of-plane interactions. Possible magnetic structures of weakly-coupled eclipsed TLA were discussed by Gekht,\cite{gekht} Chubukov\cite{chubukov} and Yamamoto.\cite{interlayer} Yet, most theoretical predictions remained unconfirmed due to the limited number of good experimental realizations of equilateral TLA.

Some of the most explored series of TLA compounds are $VX_2$ and $ABX_3$, where $A$ = Cs, Rb, $B$ = Ni, Mn, Cu, and $X$ = Cl, Br, I; as well as $ACrO_2$, with $A$ = Li, Cu, Ag, or Pd.\cite{frustration} Unfortunately, these systems either have interplanar interactions that are larger than the intraplanar ones, or too strong intraplanar nearest-neighbor coupling that makes the field-induced transitions inaccessible to the currently available magnetic-fields at the neutron scattering facilities. Improvements in sample synthesis techniques allowed in the recent years the discovery of new triangular systems with transitions at accessible magnetic fields. Detailed temperature-magnetic-field ($T-H$) phase diagrams have been reported for several $S$ = 5/2 TLA systems with planar (XY) anisotropy, RbFe(MoO$_4$)$_2$,\cite{RbFeMoO_1,RbFeMoO_2,RbFeMoO_3,RbFeMoO_4} and RbAg$_2$Fe[VO$_4$]$_2$,~\cite{RbAgFeVO} or with weak easy-axis (Ising) anisotropy,  Rb$_4$Mn(MoO$_4$)$_3$.\cite{Rb4MnMoO4_1,Rb4MnMoO4_2} Another exciting class of TLA that has recently emerged is that of 6H-perovskites Ba$_3$ $M^{\prime}M^{\prime\prime}_2$ O$_9$ with $M^{\prime}$ = Ni, Co, Mn, and $M^{\prime\prime}$= Nb, Sb or Ta.\cite{Ba3MSb2O9,Ba3NiNb2O9,Ba3MnNb2O9,Ba3CoNb2O9_1,Ba3CoSb2O9_1,Ba3CoSb2O9_2,Ba3CoSb2O9_3,
Susuki15,Ba3CoNb2O9_2,Ito17,Kamiya18,Ba3CoSb2O9_4,Rawl17,Ba3CoTa2O9} These compounds can possess either easy-plane or easy-axis anisotropies and adopt at low temperatures the expected planar 120$^\circ$ magnetic structure. For all aforementioned systems experimental evidences for the three predicted field-induced states, the ``Y'', the one-third magnetization plateau $uud$, and the ``2:1'' canted configuration, have been found. However, the lack of sufficiently large single crystal samples required for detailed neutron scattering studies has, in most of the cases, hindered the understanding of the impact of interlayer coupling in stabilization of field-induced ordered phases.

\begin{figure}[tbp]
\includegraphics[width=3.5in]{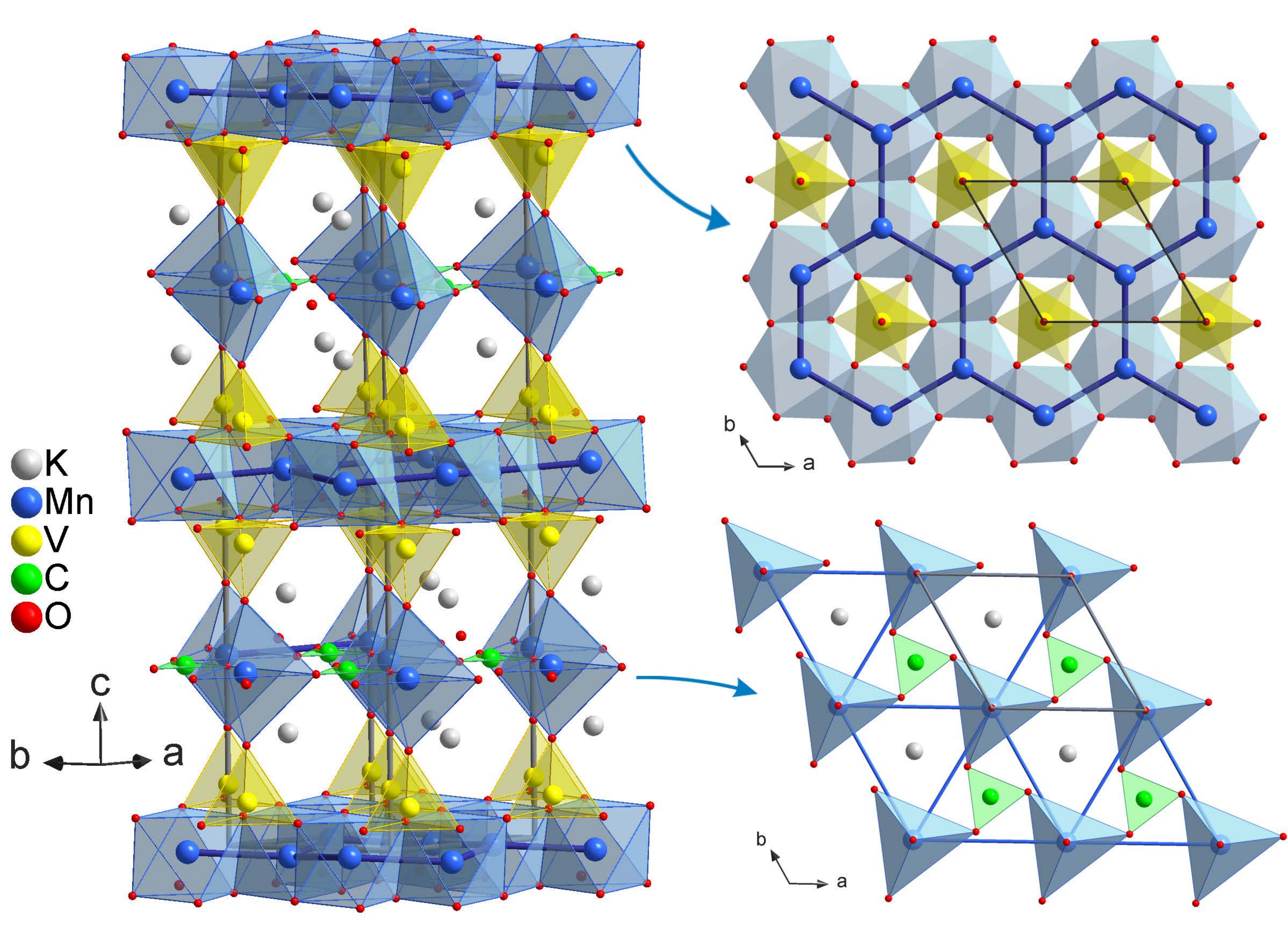}
\caption{\label{structure}(Color online) Polyhedral view of K$_{2}$Mn$_{3}$(VO$_{4}$)$_{2}$CO$_{3}$ crystal structure. The structure consists of alternate stacking of two types of layered subsystems: a honeycomb layer made of edge-sharing MnO$_6$ octahedra, and a triangular layer made of MnO$_5$ trigonal bipyramids linked by CO$_3$ triangles. Projections along $c$-axis of the two distinct layers are shown in the right panel.}
\end{figure}

The vanadate - carbonate system K$_{2}$Mn$_{3}$(VO$_{4}$)$_{2}$CO$_{3}$ has been recently identified as a very promising prototype for studying magnetic frustration.~\cite{Yakubo,Sama} Its structure, shown in Fig.~\ref{structure}, consists of alternately stacked triangular and honeycomb magnetic layers. Previous macroscopic measurements indicated complex physical properties with a presumed Jahn - Teller transformation at about 80 K and two successive magnetic phase transitions, at about 3 K and 2 K, into a weakly ferromagnetic ground state. It was inferred that divalent Mn is present in a high-spin state ($S$ = 5/2) in the octahedral environment of the honeycomb layer and a low-spin state ($S$ = 1/2) in the trigonal bipyramidal coordination of the Mn$^{2+}$ ions occupying the triangular layer.~\cite{Yakubo} The low-temperature magnetic transitions were attributed to the ordering of the $S$ = 5/2 ions of the honeycomb lattice , while the $S$ = 1/2 triangular layers were thought to remain paramagnetic. Subsequent first-principles density functional theory-based analysis showed that, contrary to the previous suggestion, both inequivalent Mn ions occupying the two layers are in high-spin $S$ = 5/2 state.~\cite{Sama} The calculations predict that both layers exhibit antiferromagnetic orders with vastly different strengths of magnetic interactions. Intrigued by the richness of the magnetic phase diagram featured by this material we have undertaken a comprehensive magnetization and neutron scattering study using high-quality single crystal samples. Our study clarifies the magnetic ground-states of two magnetic layers, and establishes that K$_{2}$Mn$_{3}$(VO$_{4}$)$_{2}$CO$_{3}$ represents an excellent candidate TLA system for studying the effect of interplanar interactions in field-induced states. Applied magnetic fields induce new magnetic superstructures characterized by various spin-stacking sequences of triangular layers. Because the observed magnetic superstructures cannot be explained by any existing theories based only on nearest-neighbor interlayer coupling, they compel a closer look at effective second-nearest-neighbor interactions mediated by fluctuations of the magnetic moments in the honeycomb sublattice.

\section{Experimental details} \label{Experiment}

Single crystals of K$_{2}$Mn$_{3}$(VO$_{4}$)$_{2}$CO$_{3}$ were grown using a high-temperature hydrothermal technique. The chemicals used in this synthesis were used as received, without further purification: Mn$_{2}$O$_{3}$ (Alfa Aesar, 98\%), V$_{2}$O$_{5}$ (Alfa Aesar, 99.6\%) and K$_{2}$CO$_{3}$ (Alfa Aesar, 99.997\%). In a typical reaction, Mn$_{2}$O$_{3}$ and V$_{2}$O$_{5}$ were mixed in a 3:2 molar ratio with 0.8 mL of 5 M K$_{2}$CO$_{3}$ mineralizer, giving approximately 0.4 g of reactants. Reactions were performed in silver ampules with a diameter of 9.5 mm, with approximately a 70\% fill of free volume. After loading the reactants and the mineralizer, the ampoules were welded and loaded in a Tuttle cold-seal style autoclave and filled with distilled water at 80\% of free volume to provide suitable counter pressure. The autoclave was heated to 580 $^\circ$C for two weeks at a typical pressure of 1.5 kbar. Brown hexagonal crystals were isolated using suction filtration. The single crystal specimens used for physical properties characterization and neutron scattering experiments were physically examined and selected under an optical microscope equipped with a polarizing light attachment. The powder sample used for the powder neutron diffraction measurements was generated by grinding crystals produces in the reaction described above.

Temperature and field-dependent magnetic measurements were carried out using a Quantum Design Magnetic Property Measurement System (MPMS). The measurements were carried out on a single crystal specimen with the crystallographic $c$-axis aligned either parallel or perpendicular to the applied magnetic field. The temperature dependence of static susceptibility ($M/H(T)$) was measured over a temperature range of 2 to 700~K for applied fields $\mu_0 H$ = 0.01 T and 1 T. The isothermal magnetization measurements were performed for fields up to 8 T. Additional isothermal magnetization curves were recorded using a vibrating sample magnetometer (VSM) in magnetic fields up to 16 T, applied along $c$-axis. Heat-capacity measurements were performed with a Physical Property Measurement System (PPMS, Quantum Design) in zero and magnetic fields up to 6 T applied either parallel and perpendicular to the crystal $c$-axis.

Neutron powder diffraction measurements were carried out using the HB2A high-resolution diffractometer at the High Flux Isotope Reactor (HFIR),\cite{hb2a} using the 2.41~{\AA} wavelength neutron beam. A powder sample with a total mass of approximately 5 grams was compacted in pellets, loaded into a cylindrical aluminium can, and placed inside a cryostat with $^{3}$He insert. Data were collected at temperatures from 150 K to 0.3 K. Single-crystal neutron diffraction measurements were performed at T = 150~K using the TOPAZ time-of-flight diffractometer at the Spallation Neutron Source (SNS). The integrated Bragg intensities were obtained using the 3-D ellipsoidal Q-space integration method and were corrected for background using the MANTID software.\cite{mantid} Data reduction including, Lorentz and absorption corrections as well as spectrum, detector efficiency, data scaling and normalization was carried out with the ANVRED3\cite{Schultz} program.

Elastic neutron scattering measurements at temperatures down to 1.6~K were performed at the fixed-incident-energy (14.6 meV) HB1A triple-axis spectrometer at the HFIR, and at CORELLI\cite{corelli} and HYSPEC\cite{hyspec} spectrometers at the SNS. Two separate crystals with the approximated dimensions 1 x 1 x 4 mm$^3$ and 4 x 3 x 1 mm$^3$ were used for collecting data under magnetic fields applied parallel and perpendicular to the $c$-axis, respectively. At CORELLI, white-beam Laue diffraction measurements were taken at 1.6~K and selected magnetic fields up to 5~T, applied along $c$ axis or [1, $\overline{1}$, 0]. For each field configuration the sample was rotated in steps of 3$^\circ$ over ranges of 60$^\circ$. MANTID software was utilized to carry out the Lorentz and spectrum corrections, as well as to merge the full volume of the scattering data.\cite{mantid} At HYSPEC, the elastic measurements were performed in applied magnetic fields up to 8~Tesla, using a fixed incident energy E$_i$ = 15~meV and a Fermi chopper frequency of 120 Hz. Measurements were concentrated over narrow range reciprocal lattice volumes, centered around the relevant magnetic reflections.

Inelastic neutron-scattering (INS) measurements were performed at the HYSPEC spectrometer, operated with the incident energies E$_i$ = 25 meV and 3.8 meV, and the Fermi chopper frequency of 360 Hz. For these measurements, multiple single crystals where coaligned along the $c$-axis, while the in-plane directions were distributed in a quasi-random manner to provide a total mass of about 0.3 g.

Structural and magnetic data refinements were carried out with the FullProf Suite program.~\cite{fullprof} Possible magnetic structures models have been explored by representation analysis using the program SARAh,\cite{sarah} and by the magnetic symmetry approach using the tools available at the Bilbao Crystallographic Server.\cite{BCS} The INS data reduction and visualization was done with the MANTID and DAVE~\cite{dave} software packages. Spin-wave calculation were performed using the linear spin wave theory with the program SpinW.\cite{SpinW}

To reproduce experimental data, we have performed classical Monte Carlo simulations using a standard metropolis sampling algorithm on the triangular lattice subsystem. The simulations were performed on finite lattice of 12 x 12 x 4 unit-cells (containing 1152 spins) with periodic boundary conditions. Starting from a completely random and disordered configuration, a spin system was annealed down to a finite temperature in finite number of intermediate temperature steps. Then, the magnetic field was increased up to 22 T with 300 intermediate field points. At each temperature/field step, the magnetization (M) and heat capacity (C) were calculated by taking the ensemble average over the 2000 samples followed by thermalization sampling with adaptive step size, while the static spin structure factor S(Q) was calculated by Fourier Transforming captured spin configurations with the frequency of 500 samples. For better statistics, all the measurements were averaged over 120 independent simulations.

\begin{table}[tbp]
\caption{\label{table1} Refined structural parameters and selected bond distances of K$_{2}$Mn$_{3}$(VO$_{4}$)$_{2}$CO$_{3}$ from single-crystal neutron diffraction data collected at T = 150~K.}
\begin{ruledtabular}
\begin{tabular}{cccccc}
Atom & Wyck. & $x$ & $y$ & $z$ & U$_{eq}$ \\[3pt]
\hline
K &$4f$& 1/3 & 2/3 & 0.6578(1) & 0.0144(4) \\[3pt]
Mn1 &$4f$& 1/3 & 2/3 & 0.0046(1) & 0.0081(3) \\[3pt]
Mn2 &$2a$& 0 & 0 & 1/4 & 0.0092(5) \\[3pt]
V &$4e$& 0 & 0 & 0.0800(4) & 0.0057 \\[3pt]
C &$2c$& 1/3 & 2/3 & 1/4 & 0.0070(3) \\[3pt]
O1 &$12i$& 0.3210(1)& 0.3098(1)& 0.0550(1)& 0.0080(1)  \\[3pt]
O2 &$4e$& 0 & 0 & 0.1552(1)& 0.0138(3) \\[3pt]
O3 &$6h$& 0.0682(1)& 0.6255(1)& 1/4 & 0.0133(3) \\[3pt]
\hline \\
\multicolumn{3}{c} {Mn1--O1 (x 3): 2.1444(6)\AA;} & \multicolumn{3}{c} { Mn2--O2 (x 2): 2.1234(8)\AA}  \\[3pt]
\multicolumn{3}{c} {Mn1--O1 (x 3): 2.1912(7)\AA;} & \multicolumn{3}{c} {Mn2--O3 (x 3): 2.1452(7)\AA}  \\[3pt]
\multicolumn{3}{c} {Mn1--Mn1 (x 3): 3.006(1)\AA;} & \multicolumn{3}{c} {Mn2--Mn2 (x 6): 5.195(1)\AA}  \\[3pt]
\hline \\
\multicolumn{6}{c}{Space group: $P6_3/m$,~$a$=$b$= 5.1959(3)\AA,~$c$ = 22.405(2)\AA}\\[3pt]
\multicolumn{6}{c} {R$_{f}$ = 0.051, $\chi^2$ = 2.04}\\[3pt]
\end{tabular}
\end{ruledtabular}
\end{table}

\section{Experimental Results and Discussion}

\subsection{Crystal structure}

The structural model of K$_{2}$Mn$_{3}$(VO$_{4}$)$_{2}$CO$_{3}$ proposed by Yakubovich \textit{et al},~\cite{Yakubo} has been confirmed by the refinements of single crystal neutron diffraction data. The refined structural parameters, such as atomic coordinates and displacement parameters along with selected bond lengths involving the magnetic Mn atoms, are given in Table~\ref{table1}. As previously described, the structure consists of two types of Mn-O layers alternating along the $c$-axis of the hexagonal unit cell. One layer consists of a honeycomb web made of edge sharing MnO$_6$ octahedra, while the second consists of MnO$_5$ trigonal bipyramids that are linked together by CO$_3$ coplanar triangle groups to form an equilateral triangular lattice. Each layer is composed by a single Mn crystallographic site: the honeycomb is built of Mn1 ions occupying $4f$ Wyckoff position of the $P6_3/m$ space group, and the triangular layer is defined by Mn2 ions located at the $2a$ Wyckoff position. The Mn1 atoms are coupled via double oxygen (O1) bridges and are spaced at about 3.006~\AA~apart. Inside the triangular layer the interatomic distance between neighboring Mn2 atoms is approximately 5.195~\AA. There are twice as many Mn1 ions in the honeycomb layer as compared to Mn2 located in the triangular layer. The Mn2 ions are located exactly on top or underneath the hollow center of the Mn1 honeycomb. Considering that there are two layers of each type per unit cell, the distance between two consecutive triangular layers is about $c$/2 = 11~\AA, while that between the honeycomb and triangular layer is $c$/4 = 5.5~\AA. The interlayer space is occupied by K$^+$ cations and V$^{5+}$O$_4$ tetrahedra that share oxygen vertices with manganese polyhedra. Due to the non-magnetic nature of V$^{5+}$ cation, the magnetic interactions between adjacent layers are expected to be subdominant compared to the intralayer interactions. A perspective view of the crystal structure and the two types of Mn-O layers are depicted in Fig.~\ref{structure}.

It is instructive to compare here the structural properties of our compound with other recently studied triangular lattice systems, where Mn$^{2+}$ magnetic ions adopt either six-fold or five-fold oxygen coordinations. For instance, the Rb$_4$Mn(MoO$_4$)$_3$ features Mn$^{2+}$O$_5$ polyhedra forming equilateral triangular lattices separated by MoO$_4$ tetrahedra, with intralayer and interlayer distances between Mn$^{2+}$ ions of 6.099~\AA~ and $c$/2= 11.856~\AA, respectively.\cite{Rb4MnMoO4_1} In Ba$_3$MnNb$_2$O$_9$ the Mn$^{2+}$ ions have a octahedral coordination and form triangular lattices with Mn-Mn intralayer distance of 5.773~\AA~ and interlayer distances $c$ = 7.0852~\AA.\cite{Ba3MnNb2O9}

The evolution with temperature of K$_{2}$Mn$_{3}$(VO$_{4}$)$_{2}$CO$_{3}$ lattice parameters has been investigated by powder neutron diffraction and single-crystal X-ray measurements. A smooth temperature dependence has been observed, suggesting that no noticeable structural change takes place down to 1.7~K. This would appear to disprove the earlier speculations in Ref.~\cite{Yakubo} of a Jahn - Teller distortion taking place at around 80 -- 100 K.

\subsection{Macroscopic magnetic behaviour}

\begin{figure}[tbp]
\includegraphics[width=3.4in]{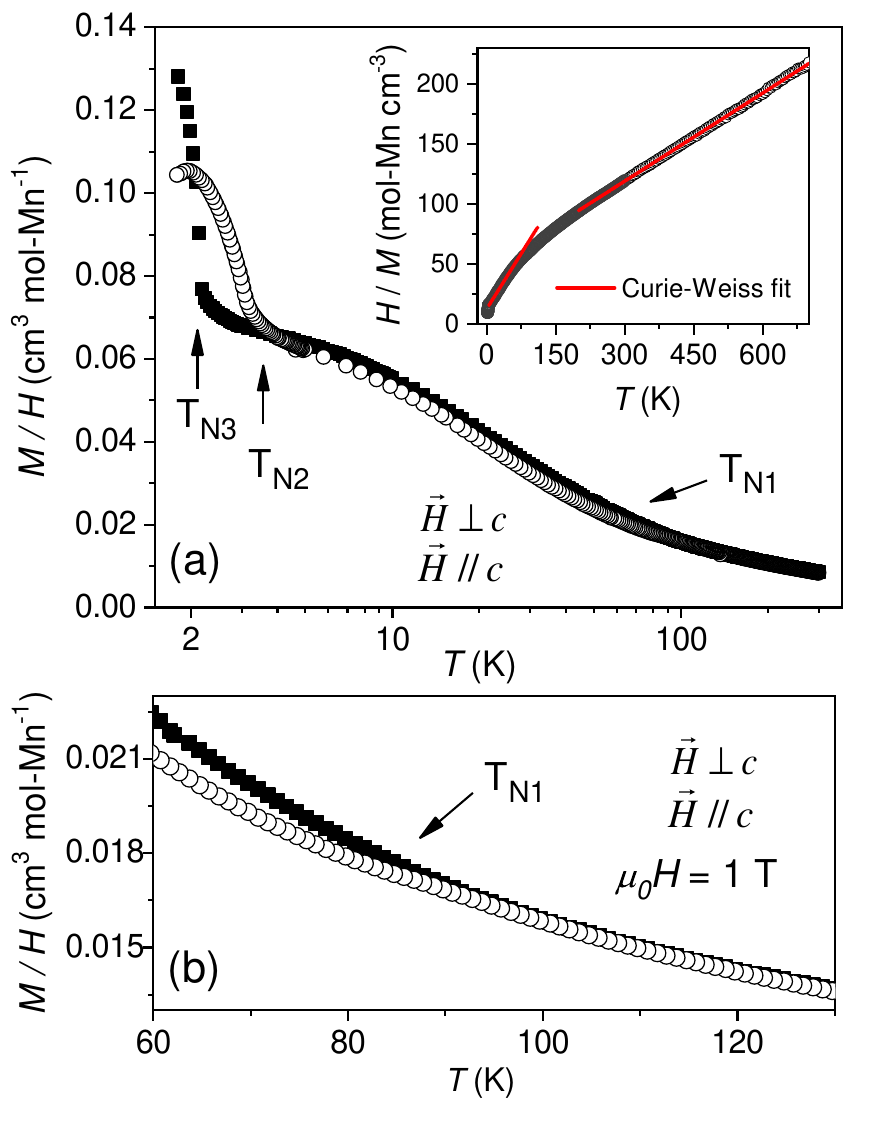}
\caption{\label{suscept} (a) Evolution of the magnetic susceptibility ($M/H$) as a function of temperature, measured in a field of 1~T applied parallel (open symbol) and perpendicular (solid symbol) to the $c$-axis. The temperature axis is given in a logarithmic scale to increase the visibility of the magnetic phase transitions. The inset shows the inverse susceptibility that features two linear regimes at 200 K $< T <$ 700 K and 10 K $< T <$ 50 K. The Curie-Weiss fits are shown as red lines. (b) Expanded view of magnetization curves around the $T_{N1}$ transition temperature. The magnetic order is  revealed by a subtle drop in the magnetization curve measured with field applied along the $c$ direction.}
\end{figure}

\begin{figure}[tbp]
\includegraphics[width=3.6in]{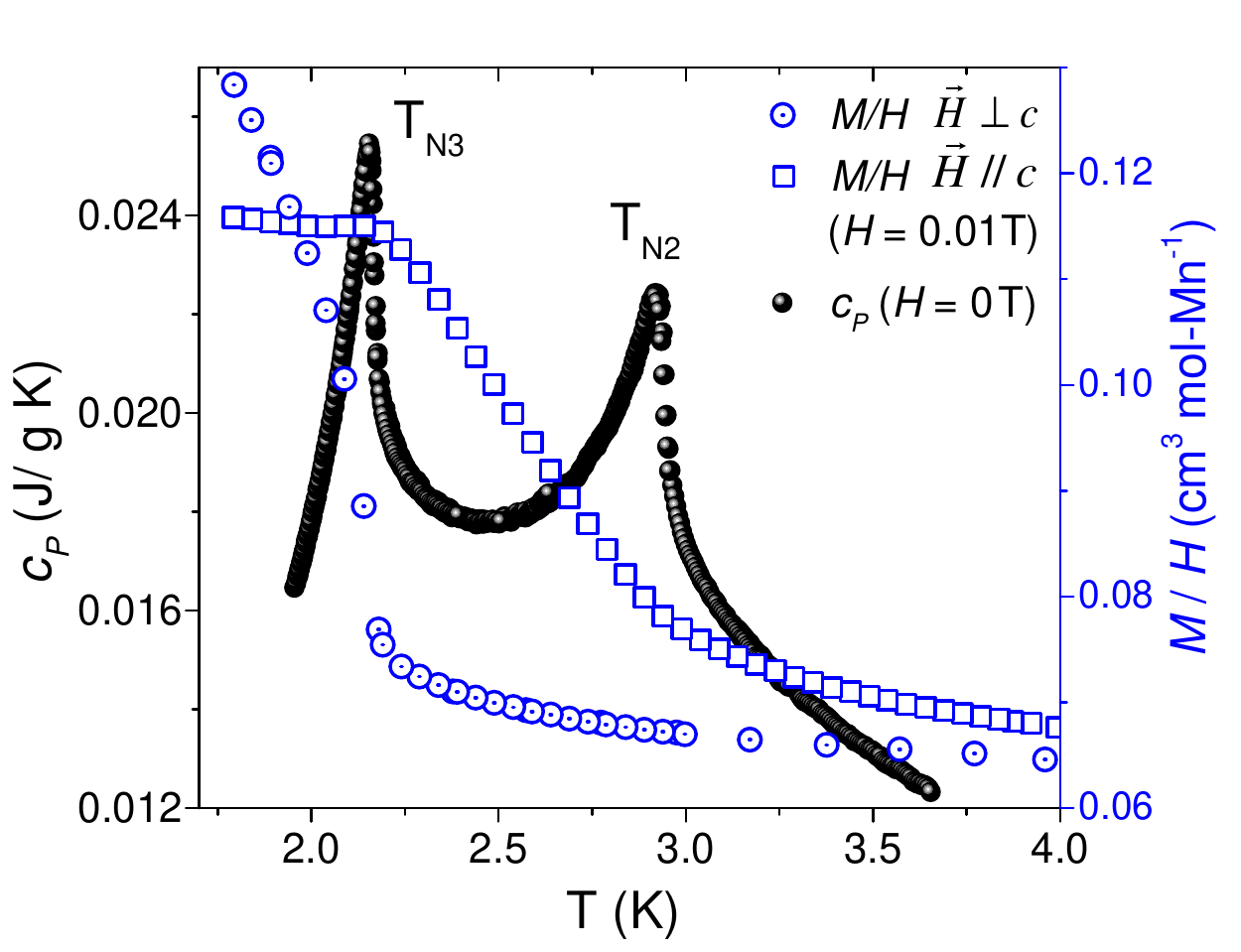}
\caption{\label{heatcap} Low-temperature region of the zero-field specific heat, and low-field static magnetic susceptibility ($M/H$) measured with $\mu_0H$ = 0.01 T for two different field orientations, $\protect\overrightarrow{H}\parallel c$ and $\protect\overrightarrow{H}\perp c$, revealing the two successive magnetic transitions at about 3 K and 2.2 K.}
\end{figure}

The temperature dependence of magnetic susceptibility ($\chi=M/H$) measured with a magnetic field applied along and perpendicular the $c$-axis is shown in Fig.~\ref{suscept}. In this figure we choose to plot the temperature axis in a logarithmic scale to increase the visibility of the magnetic phase transitions that occur over a large temperature interval. A first magnetic transition near $T_{N1} \approx$ 85 K is revealed by a subtle drop in the susceptibility curve measured with field applied along $c$ direction. A second magnetic order transition is discernible as a rise of the $\overrightarrow{H}\parallel c$ susceptibility at $T_{N2} \approx$ 3~K. This is immediately followed by a third transition, which appears as a kink in the susceptibility curve at $T_{N3} \approx$ 2.2~K. The large difference between the $T_{N1}$ and $T_{N2}$ ordering temperatures denotes that the honeycomb and triangular magnetic layers possess magnetic interactions of completely different energy scales and act as nearly independent magnetic sublattices.

The low-temperature region of the zero-field heat capacity data, together with magnetic susceptibility data corresponding to two different magnetic field orientations, $\overrightarrow{H}$ $\parallel$ $c$ and $\overrightarrow{H}$ $\perp$ $c$, are displayed in Fig.~\ref{heatcap}. The two successive magnetic transitions at approximately 3 K and 2.2 K can be clearly seen in heat capacity data ($c_P$) as two distinguishable lambda-shaped peaks. One can also notice in Fig.~\ref{heatcap} that the magnetic susceptibility exhibits significant anisotropic behavior. When the field is applied parallel to the $c$-axis the susceptibility undergoes changes at both $T_{N2}$ and $T_{N3}$ transitions. However, when the field is applied perpendicular to the $c$-axis, only the lower transition near 2.2~K ($T_{N3}$) is visible. This behaviour is indicative of a spin-canting taking place at the lowest temperature.

The inverse magnetic susceptibility is shown in the inset of Fig.~\ref{suscept} along with fits using the Curie-Weiss model ($\chi$ = $\chi_0$ + C/($T$ - $\Theta_{CW}$)). The $1/\chi$ data shows two linear regimes: one for the temperature range 200 to 700 K where all spins are paramagnetic, and a second at lower temperatures ranging from 10 to 50 K. For the high temperature range, the Curie-Weiss fit yields a Curie constant C = 4.44 cm$^3$/mol-Mn/K, a Weiss temperature of -215 K, and a temperature-independent term $\chi_0$ = -0.00027 cm$^3$/mol-Mn. The negative experimental value of $\chi_0$ can be attributed to the diamagnetic background from the sample holder, as the core diamagnetism correction is estimated to be about an order of magnitude smaller that the obtained value. The negative Weiss temperatures indicates dominant antiferromagnetic interactions. The obtained effective moment 5.95~$\mu_B$/Mn is very close to that expected for $S$ = 5/2 Mn$^{2+}$, $g \sqrt{S(S+1)}$ = 5.91~$\mu_B$/Mn, suggesting that Mn ions have the same spin state in both honeycomb and the triangular planes. A Curie-Weiss fit performed using a constrained $\chi_0$ = 0, yields a slightly lower effective moment 5.72 $\mu_B$/Mn and $\Theta_{CW}$ = -188 K. Note that previous study\cite{Yakubo} reported much smaller values for both the effective moment (2.75~$\mu_B$/Mn) and the Curie-Weiss temperature (−114 K), which were likely caused by an overestimation of the temperature-independent susceptibility term. Although the temperature interval of the second Curie-Weiss regime, 10 K $<$ T $<$ 50 K, is too narrow for extracting definite information, the Curie-Weiss fit yields a Weiss temperature of about -19 K and Curie constant 1.62 cm$^3$/mol-Mn/K. One could remark that this Curie constant represents about 36\% of the value obtained for the high temperatures and it correlates reasonably well with the 1/3 fraction of spins located in the triangular layer. This suggests that the honeycomb and triangular layers act as nearly independent magnetic sublattices.

\begin{figure}[]
\includegraphics[width=3.4in]{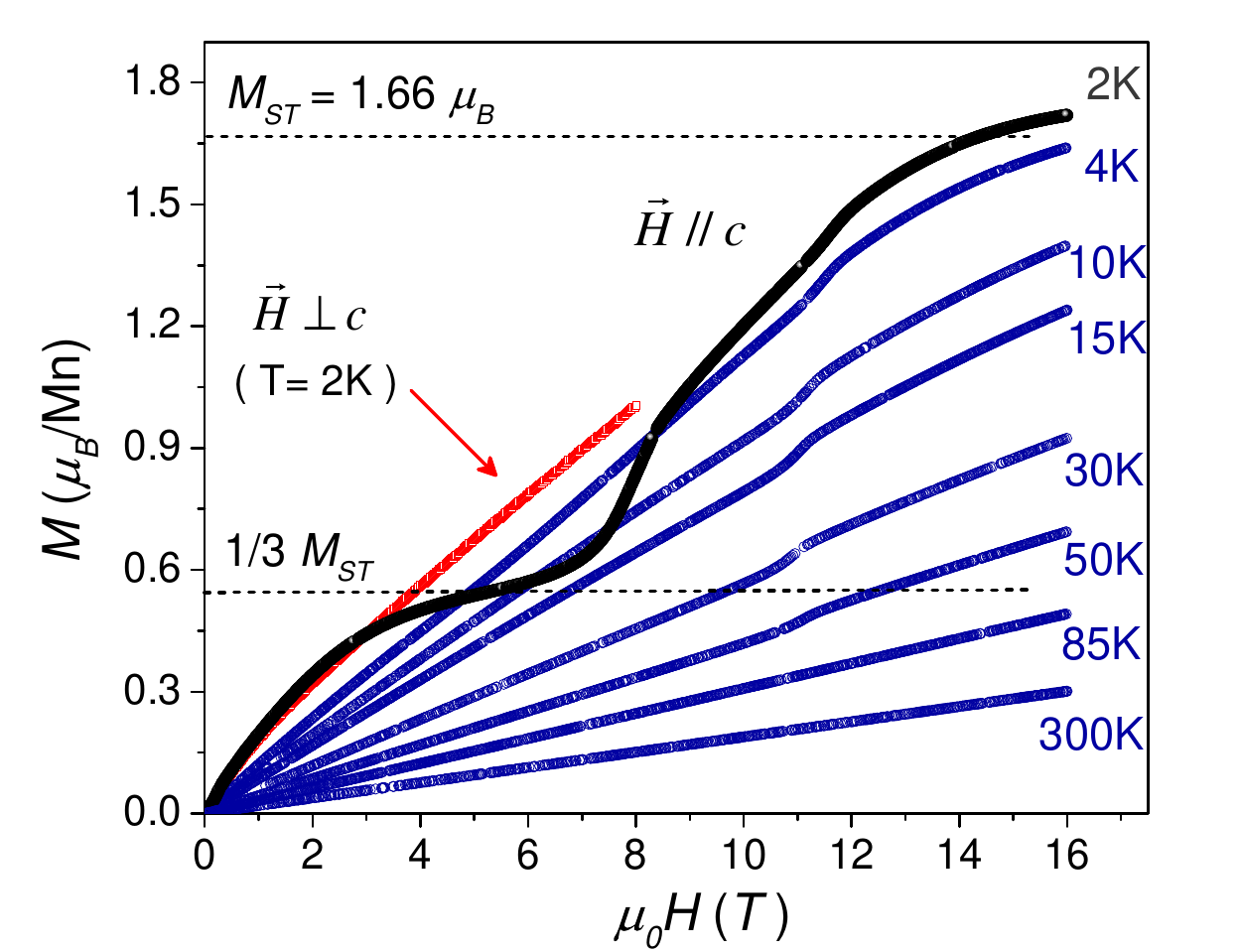}
\caption{\label{MH} (Color online) Isothermal magnetization ($M$ vs $H$) curves measured at selected temperatures ranging from 2 K to 300 K for magnetic fields applied along $c$-axis (black and blue curves). Magnetization curve measured at 2 K for magnetic fields applied perpendicular to the $c$-axis is shown in red color and indicated by arrow. The magnetization values corresponding to the plateau state (1/3 $M_{ST}$) and saturation $M_{ST}$ of the triangular magnetic sublattice are indicated by a dashed line. As described in the text, the triangular layer contains only one third of the total magnetic ions of the system.}
\end{figure}

\begin{figure}[]
\includegraphics[width=3.4in]{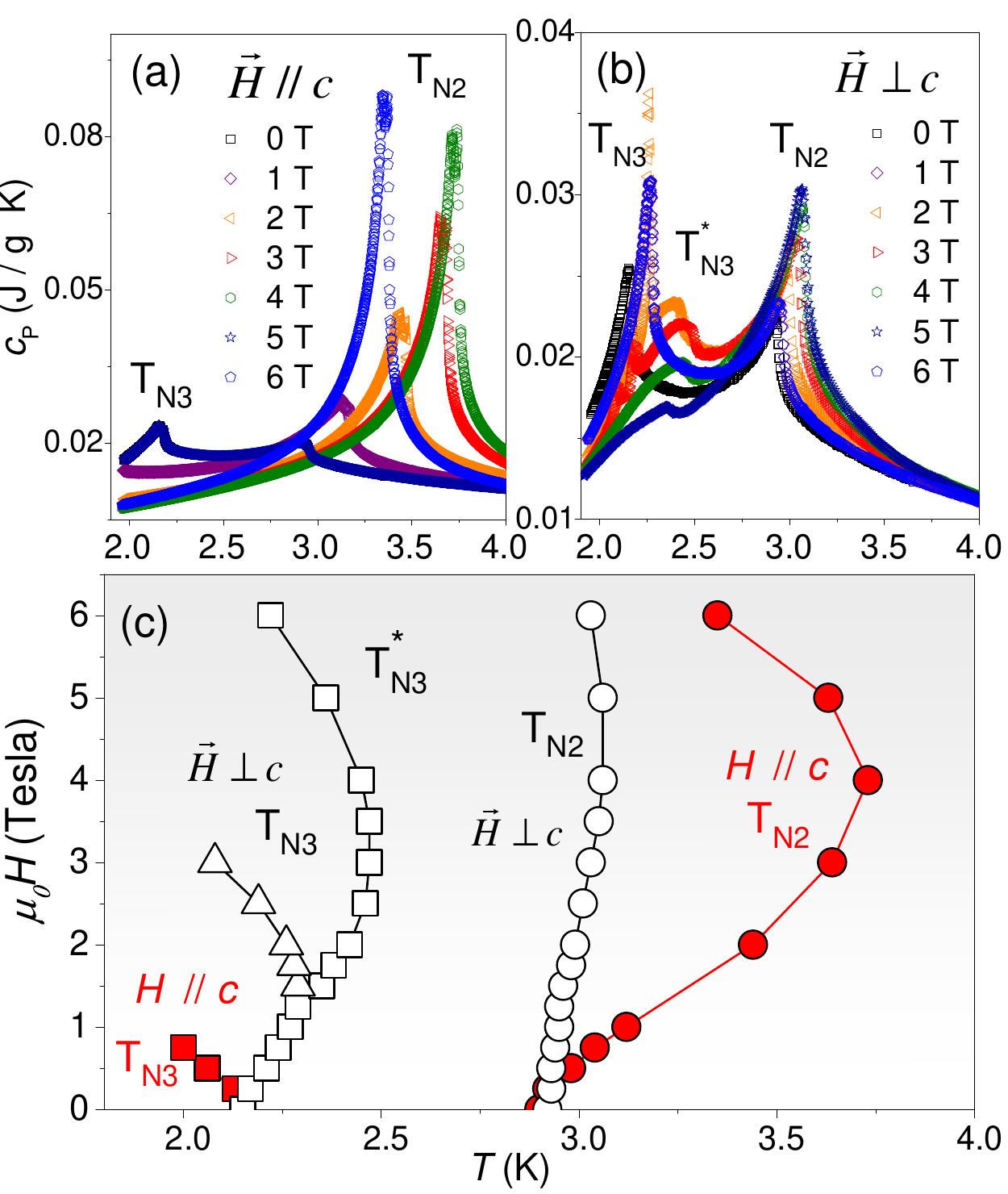}
\caption{\label{heatcapfield} (Color online) Low-temperature region of the heat-capacity data measured in applied magnetic fields oriented parallel (a) and perpendicular (b) to the $c$-axis direction. (c) The partial $T -\mu_0H$ magnetic phase diagram deduced from heat-capacity data. For $\protect\overrightarrow{H}\parallel c$, denoted by solid symbols, the lowest transition ($T_{N3}$) shifts to lower temperatures with increasing the magnetic field, while the $T_{N2}$ transition displays a dome-like shape. For $\protect\overrightarrow{H}\perp c$, represented by open symbols, the $T_{N3}$ transition evolves into two distinct transitions ($T_{N3}$ and $T^{*}_{N3}$) while the $T_{N2}$ transition increases slightly.}
\end{figure}

The isothermal magnetization curves measured at selected temperatures ranging from 2 K to 300 K are shown in Fig.~\ref{MH}. Magnetization data was taken for magnetic fields up to 16 T applied along $c$-axis, and up to 8 T for fields applied perpendicular to the $c$-axis. In agreement with the previous report,\cite{Yakubo} the magnetization curve measured with $\overrightarrow{H}$ $\parallel$ $c$ at $T$ = 2 K exhibits a plateau at about 4.5 Tesla, followed by a sharp upturn at approximately 7 T. Another step-like transition is observed near 11 T, and then a tendency towards saturation as the field approaches 16 T. The magnetization value of the plateau state ($\approx$ 0.55 $\mu_B$/Mn) appears to correspond to about 1/9 of total saturation value of Mn$^{2+}$ moments (M$_S$ = $g$S = 5~$\mu_B$), or 1/3 of saturation value of Mn$^{2+}$ moments located in the triangular layers. The magnetization plateau value, labeled as 1/3 M$_{ST}$, is marked by a dashed line in Fig.~\ref{MH}. We note that the saturation of magnetization of the triangular layers, M$_{ST}$ = 1.66 $\mu_B$/Mn, is found to be reached at a magnetic field of about 14.3 Tesla. At T = 4 K, the plateau-like state in the $\mu_0\overrightarrow{H}$ $\parallel$ $c$ magnetization curve is no longer present, suggesting that this is related to the low-temperature ordering states observed below 3 K ($T_{N2}$). On the other hand, the 11 T step-like transition persists up to 50 K, appearing to be related to the higher-temperature ordered state that emerges at $T_{N1}$ $\approx$ 85 K. There is no apparent rational number correlation between the magnetization value corresponding to this transition and the total saturation value M$_S$. In contrast to the magnetization curve measured with $\overrightarrow{H}$ $\parallel$ $c$, the curve measured with $\overrightarrow{H}$ $\perp$ $c$ shows a smooth increase with increasing the magnetic field up to the highest measured value of 8 T.

Heat capacity measurements in applied magnetic fields have been carried out to construct the $T-H$ phase diagram around the two low-temperature magnetic transitions. The results are summarized in Fig.~\ref{heatcapfield}. For the field oriented along the $c$-axis direction, the $T_{N3}$ magnetic transition shifts quickly towards lower temperatures with increasing the magnetic field, while the intermediate transition $T_{N2}$ extends first to higher temperatures and then diminishes to define a dome-like shape with the tip at approximately $\mu_0 H$ = 4 T and 3.75 K. The transition points for $\overrightarrow{H}$ $\parallel$ $c$ are represented by filled symbols in Fig.~\ref{heatcapfield}(c). It is also interesting that the heat capacity peak corresponding to the intermediate transition displays a dramatic increase in intensity as the field increases to 4 T, after which it remains relatively flat. This $\overrightarrow{H}$ $\parallel$ $c$ phase diagram is reminiscent of that of a triangular lattice antiferromagnet with weak easy-axis anisotropy, and it is strikingly similar to that observed for Rb$_4$Mn(MoO$_4$)$_3$\cite{Rb4MnMoO4_1} and Ba$_3$MnNb$_2$O$_9$~\cite{Ba3MnNb2O9}.

Upon applying the magnetic field perpendicular to $c$-axis the heat capacity peak corresponding to the $T_{N2}$ transition shifts only slightly to higher temperatures, whereas the peak denoting the $T_{N3}$ transition splits into two components that display dome-shaped profiles as a function of magnetic field. The higher-temperature component that emerges from the zero-field heat capacity peak, labeled as $T_{N3}^*$ in Fig.~\ref{heatcapfield}, is much broader and reduced in amplitude. The phase diagram for this field direction is unexpectedly more complicated than that observed in other triangular lattice antiferromagnets.

\subsection{Zero-field magnetic order}

\subsubsection{Magnetic order of the honeycomb sublattice}

\begin{figure}[tbp]
\includegraphics[width=3.4in]{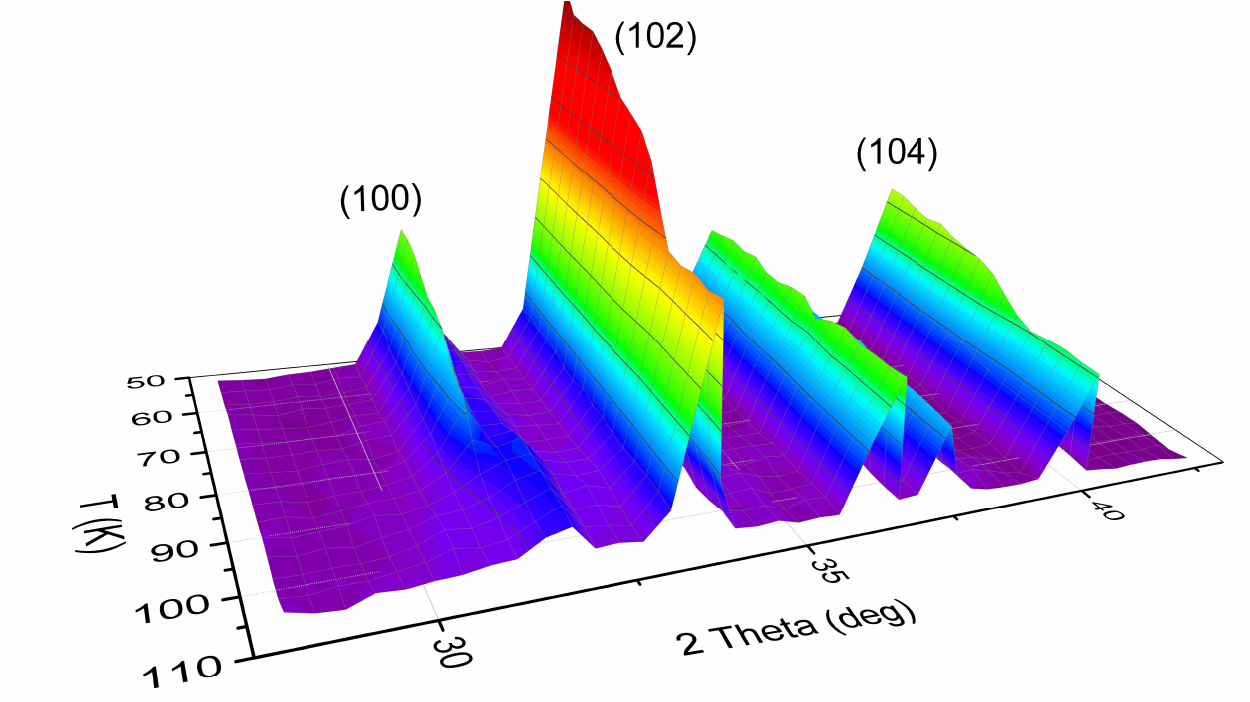}
\caption{\label{powderdiff} (Color online) Contour plot of the evolution of the neutron powder diffraction patterns recorded on cooling from 110 K to 50 K. The data reveals the appearance of magnetic Bragg reflections of the type(1, 0, $L$=2$n$) corresponding to the long-range magnetic ordering of the Mn1 ions occupying the honeycomb layers.}
\end{figure}

\begin{figure}[btp]
\includegraphics[width=3.6in]{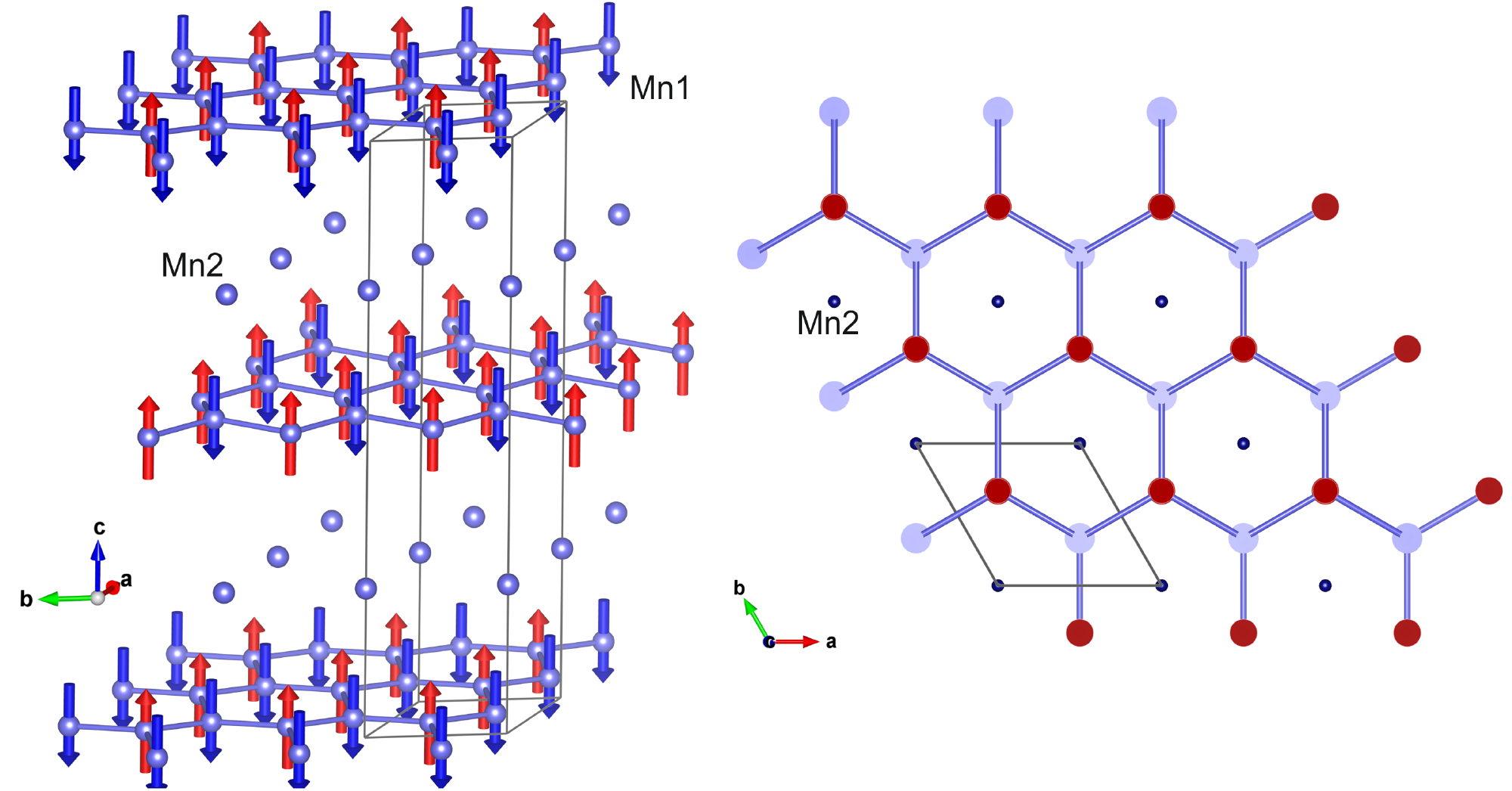}
\caption{\label{honeymagstr} (Color online) Magnetic structure for the temperature range 3 K $\leqslant T \leqslant$ 85 K, defined by a Ne\'{e}l-type antiferromagnetic arrangement of Mn1 moments in the honeycomb layer, while the Mn2 in triangular layers remain paramagnetic. The ordered moments are aligned along the $c$ direction. The successive antiferromagnetic planes are stacked ferromagnetically. The right panel display a projection of the structure along $c$-axis, emphasizing the location of paramagnetic Mn2 site on top or underneath the zero molecular field created by the six surrounding Mn1 ordered moments.}
\end{figure}

Powder and single crystal neutron diffraction data collected below $T_{N1}$ reveal additional scattering at low angles reflections, of the type(1, 0, $L$=2$n$). The evolution of the powder diffraction pattern across this first magnetic transition is shown in Fig.~\ref{powderdiff}. Magnetic structures models compatible with the space group $P6_3/m$ and the propagation vector $\textbf{k}$=(0, 0, 0) have been explored using both the magnetic symmetry approach, using MAXMAGN program,~\cite{BCS} and the representation analysis with the program SARAH.~\cite{sarah}. Among the four possible maximal magnetic space groups, the $P6_3^\prime/m$ (\#176.145) is the only one that fits well all observed magnetic intensities. The magnetic structure at intermediate temperatures, 3 K $\leqslant$ T $\leqslant$ 85 K, consists on a Ne\'{e}l-type antiferromagnetic order characterized by antiparallel alignment of nearest-neighbor Mn1 moments in the honeycomb layer. The magnetic moments at the Mn2 sites of the triangular layer remain disordered. The Mn1 moments are oriented parallel to the $c$-axis and the antiferromagnetic honeycomb layers are stacked ferromagnetically along the $c$-axis direction. The refined amplitude of the static moment is 2.9(1)$\mu_B$ at 50~K , and it converges to 5.0(1)$\mu_B$ at 1.7 K. The magnetic structure is depicted in Fig.~\ref{honeymagstr}. The magnetic moment orientation for each atomic position is explicitly given in Table~\ref{table2}. It is worth noting that the N\'{e}el-type AFM ground state of the honeycomb lattice is susceptible to undergo a spin-flop transition for a magnetic field applied parallel to $c$-axis.\cite{Li2MnO3} Thus, the step-like anomaly observed at about 11 T in the isothermal magnetization measurements can be interpreted as a spin-flop transition.

\begin{table}[tbp]
\caption{\label{table2} Magnetic structure configuration of K$_{2}$Mn$_{3}$(VO$_{4}$)$_{2}$CO$_{3}$ for the temperature range 3 K $\leqslant$ T $\leqslant$ 85 K, described by $\textbf{k}$=(0, 0, 0) and the magnetic space group $P6_3^\prime/m$.}
\begin{ruledtabular}
\begin{tabular}{rcc}
\vspace{.1in}
Atom & ( $x$, $y$, $z$ )&( $m_a$, $m_b$, $m_c$ )\\[3pt]
\hline
Mn1 &( 1/3,~2/3,~0.00460(5) )&( 0,~0,~$m_c$ )\\[3pt]
    &( 1/3,~2/3,~0.4954(5) )&( 0,~0,~$m_c$ )\\[3pt]
    &( 2/3,~1/3,~-0.00460(5) )&( 0,~0,~$-m_c$ )\\[3pt]
    &( 2/3,~1/3,~0.50460(5) )&( 0,~0,~$-m_c$  )\\[3pt]
Mn2 &( 0,~0,~1/4 ) &( 0,~0,~0 )\\[3pt]
    &( 0,~0,~3/4 ) &( 0,~0,~0 ) \\[3pt]
\end{tabular}
\end{ruledtabular}
\end{table}

\begin{figure}[btp]
\includegraphics[width=3.4in]{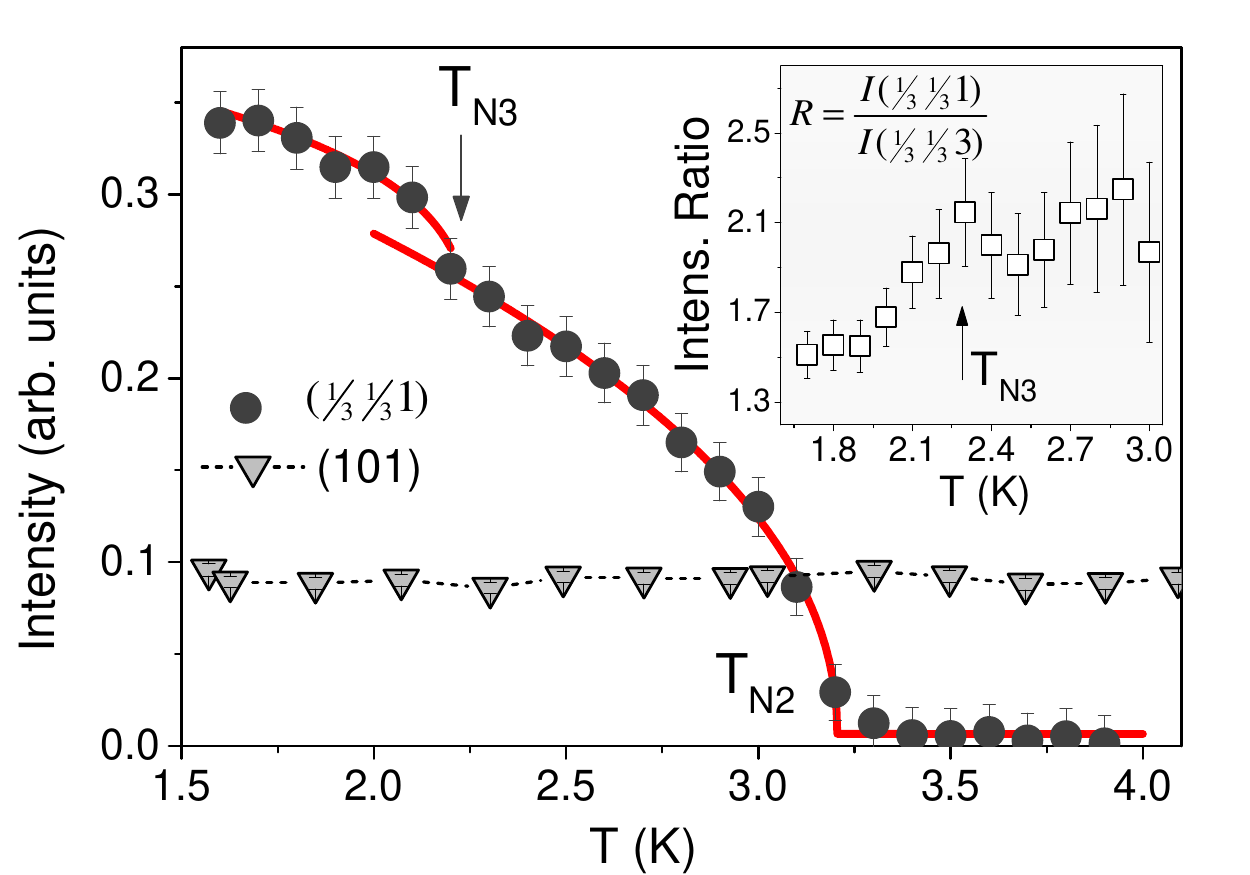}
\caption{\label{orderparam} (Color online) Temperature dependence of ($\frac{1}{3}$, $\frac{1}{3}$, 1) magnetic peak intensity revealing the long-range magnetic order of Mn2 ions on the triangular layer at $T_{N2}$ $\approx$ 3.2~K , followed by a spin reorientation at $T_{N3}$ $\approx$ 2.2~K. The absence of magnetic scattering contribution to the (1,0,1) reflection across the two low-temperature transitions suggests that the intermediate magnetic order consists of a collinear $up-up-down$ structure with modulated-amplitude (see text for details). The insert displays the temperature dependence of the  ratio of the intensities ($\frac{1}{3}$, $\frac{1}{3}$, 1) and ($\frac{1}{3}$, $\frac{1}{3}$, 3). The larger gain in intensity of ($\frac{1}{3}$, $\frac{1}{3}$, 3) below $T_{N3}$ demonstrates that magnetic moments are rotating away from the $c$-axis to produce a ``Y''-type spin structure.}
\end{figure}

\begin{figure}[]
\centering
\subfigure[]{\includegraphics[width=3.5in]{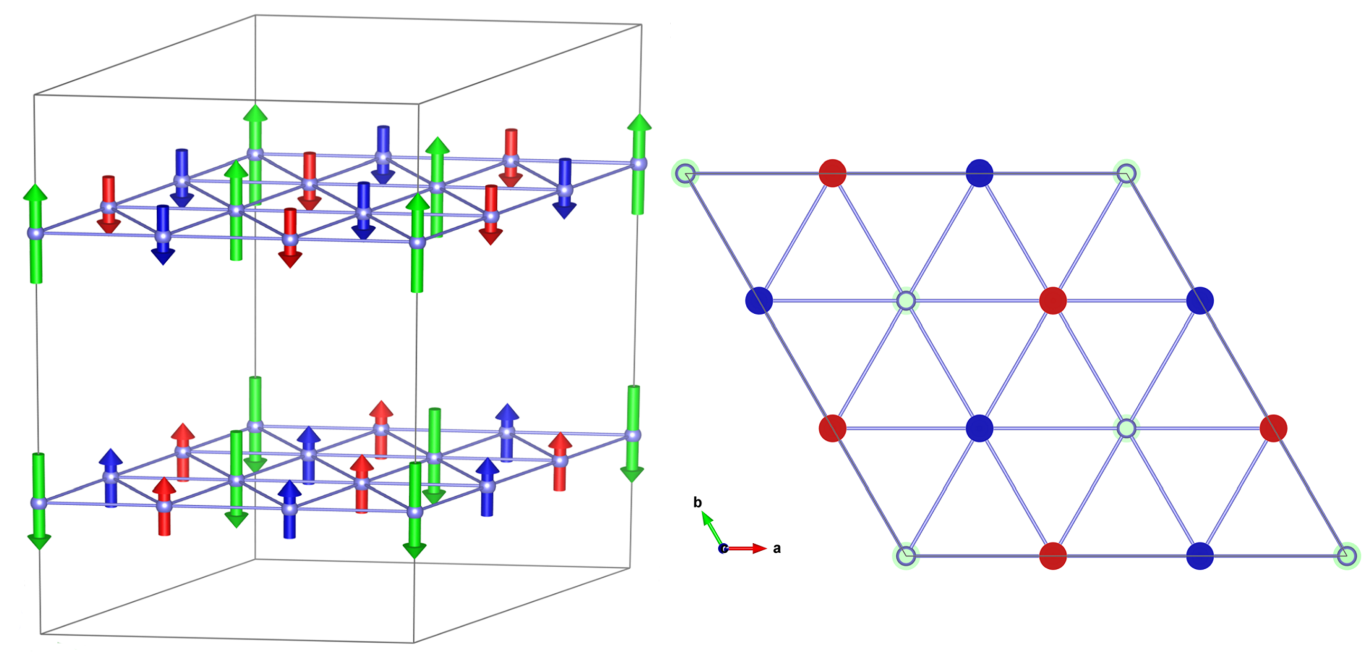}}
\subfigure[]{\includegraphics[width=3.5in]{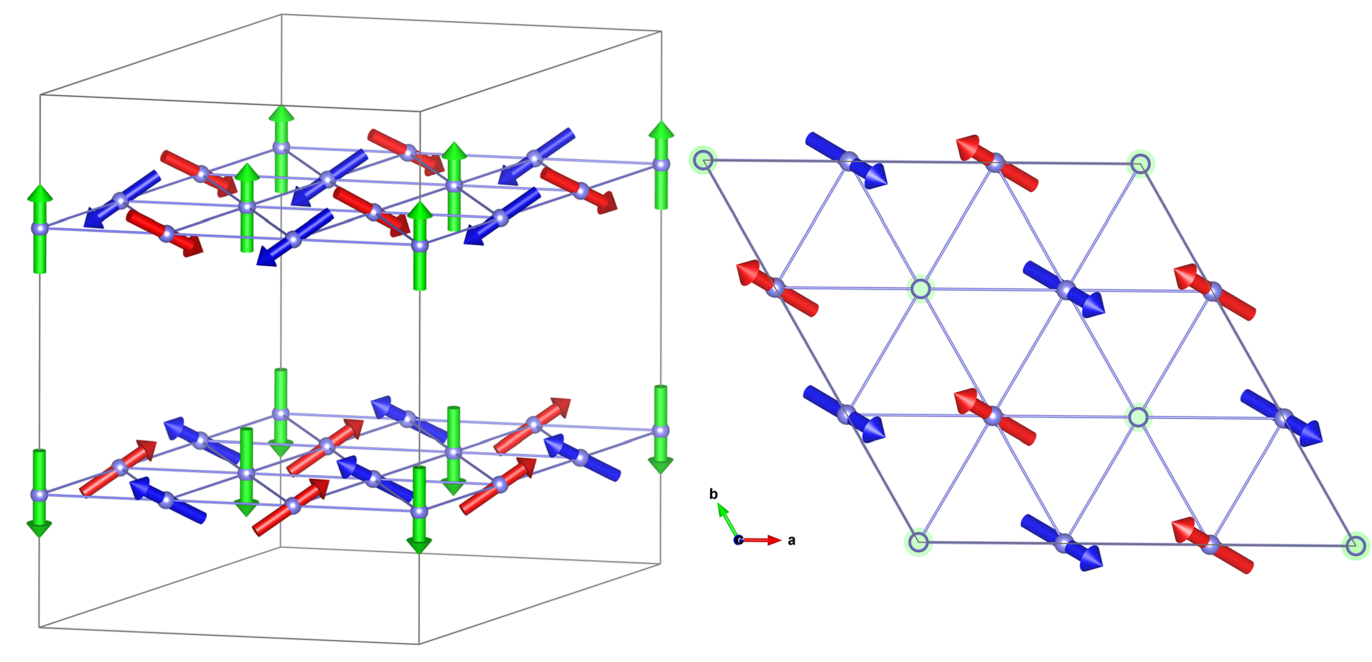}}
\caption{\label{magstruct} (Color online) (a) Three-sublattice static magnetic order of Mn2 atoms (inside the triangular layer) in the intermediate temperature 2.2 K $\leqslant T \leqslant$ 3 K. The magnetic moments are aligned parallel to the $c$-axis in an $up-up-down$ configuration. The moment amplitude follows the $\textbf{k}$=($\frac{1}{3}$, $\frac{1}{3}$, 0) modulation, $m_i=m_0 \Re[cos(2\pi\textbf{k}\cdot\textbf{r}_i]$, resulting in a fully compensated magnetization inside the plane. Successive triangular layers are stacked in an antiparallel manner. The honeycomb layers maintain the AFM structure shown in Fig.~\ref{honeymagstr} but are omitted for clarity. A view along the $c$-axis of a three-sublattice collinear magnetic order is shown in the right panel. (b) Three-sublattice magnetic structure below 2.2 K. Two of Mn2 atoms develop in-plane ($ab$ ordered spin components to form a planar canted \textbf{Y}-type structure, where moments are rotated by about 120$^\circ$ between neighboring sites. The structure remains bilayer with the moments of adjacent layers being antiparallel to each other. The right panel of the figure displays the projection of the spin structure on the $ab$ basal plane.}
\end{figure}

\subsubsection{Magnetic order of the triangular sublattice}

A new set of magnetic reflections with  propagation vector \textbf{k} = ($\frac{1}{3}$, $\frac{1}{3}$, 0) appear upon cooling below approximately 3.2 K. The temperature evolution  of the ($\frac{1}{3}$, $\frac{1}{3}$, 1) peak intensity is shown in Fig.~\ref{orderparam}. The observed transition point agrees well with T$_{N2}$ obtained from macroscopic measurements. Furthermore, a kink in the order parameter curve is observed near 2.2~K, that matches the position of the second peak at $T_{N3}$ in the heat capacity data. No change in intensity of the $\textbf{k}$ = (0,~0,~0) -type magnetic peaks associated with the ordering of Mn1 honeycomb sublattice was observed to occur at these low temperatures. The spin reorientation suggested by the static susceptibility data to take place at T$_{N3}$ is best captured by following the temperature dependence of the ratio between ($\frac{1}{3}$, $\frac{1}{3}$, 1) and ($\frac{1}{3}$, $\frac{1}{3}$, 3) peak intensities, shown in the insert of Fig.~\ref{orderparam}. It is noticeable that the ($\frac{1}{3}$, $\frac{1}{3}$, 3) exhibits a larger gain in intensity below 2.2 K.  Considering that only the moment component perpendicular to the momentum transfer \textbf{Q} contributes to the magnetic intensities, the abrupt change in the relative intensity gain indicates that some magnetic moments are rotating away from the $c$-axis. A rough estimate of the canting angle $\Theta$ = 54(5)$^\circ$ is obtained from the intensities ratio at $T$ = 1.7 K.

\begin{table}[btp]
\caption{\label{table3} Magnetic spin configuration of Mn2 site for the temperature range 2.2 K $\leqslant$ T $\leqslant$ 3 K, and below 2.2 K, described by $\textbf{k}$=($\frac{1}{3}$, $\frac{1}{3}$, 0) and magnetic space groups $P6_3^\prime/m$ and $P2_1^\prime$, respectively. The atomic coordinates ($x^\prime$, $y^\prime$, $z^\prime$) are defined for the 3 x 3 x 1 magnetic super-cell. In the expanded lattice there are three non-equivalent magnetic sites and moments projections in the $ab$ plane and $c$-direction  follow the modulation imposed by $\textbf{k}$ wave-vector (as described in the text). We constrained the moment directions to form an ideal 120$^\circ$ structure with the in-plane projection $m_{ab}\parallel$[1,1,0] direction. The refined magnitude of the static moment $m_0$ at 1.7 K is 3.7(2)$\mu_B$.}
\begin{ruledtabular}
\begin{tabular}{rccc}

& & 2.2 K $\leqslant$ T $\leqslant$ 3 K  & T $<$ 2.2 K  \\[3pt]
& & $P6_3^\prime/m$  & $P2_1^\prime$ \\[3pt]
Atom & ($x^\prime$ , $y^\prime$ , $z^\prime$ ) & ($m_a$,$m_b$,$m_c$)  & ($m_a$,$m_b$,$m_c$) \\[5pt]
\hline

Mn2$_1$ & (0,0,1/4) & (0,0,$m_0/2$) & ($m_0/2$,$-m_0/2$,$m_0/2$) \\[3pt]
      & (0,2/3,3/4) & (0,0,$-m_0/2$)& ($m_0/2$,$-m_0/2$,$-m_0/2$)  \\[3pt]
      & (1/3,1/3,3/4) & (0,0,~$-m_0/2$)& ($m_0/2$,$-m_0/2$,$-m_0/2$) \\[3pt]
      & (1/3,2/3,1/4) & (0,0,$m_0/2$)& ($m_0/2$,$-m_0/2$,$m_0/2$) \\[3pt]
      & (2/3,0,3/4) & (0,0,$-m_0/2$)& ($m_0/2$,$-m_0/2$,$-m_0/2$)  \\[3pt]
      & (2/3,1/3,1/4) & (0,0,$m_0/2$)& ($m_0/2$,$-m_0/2$,$m_0/2$)  \\[3pt]
Mn2$_2$ & (0,0,3/4) & (0,0,$-m_0/2$) & ($-m_0/2$,$m_0/2$,$-m_0/2$) \\[3pt]
      & (0,2/3,1/4) & (0,0,$m_0/2$) &($-m_0/2$,$m_0/2$,$m_0/2$)  \\[3pt]
      & (1/3,1/3,1/4) & (0,0,$m_0/2$) &($-m_0/2$,$m_0/2$,$m_0/2$)  \\[3pt]
      & (1/3,2/3,3/4) & (0,0,$-m_0/2$) &($-m_0/2$,$m_0/2$,$-m_0/2$)  \\[3pt]
      & (2/3,0,1/4) & (0,0,$m_0/2$) &($-m_0/2$,$m_0/2$,$m_0/2$)  \\[3pt]
      & (2/3,1/3,3/4) & (0,0,$-m_0/2$) &($-m_0/2$,$m_0/2$,$-m_0/2$)  \\[3pt]
Mn2$_3$ & (0,1/3,1/4) & (0,0,$-m_0$) &(0,0,$-m_0$) \\[3pt]
      & (0,1/3,3/4) & (0,0,$m_0$) & (0,0,$m_0$)  \\[3pt]
      & (1/3,0,1/4) & (0,0,$-m_0$) & (0,0,$-m_0$)  \\[3pt]
      & (1/3,0,3/4) & (0,0,$m_0$) & (0,0,$m_0$)  \\[3pt]
      & (2/3,2/3,1/4) & (0,0,$-m_0$) & (0,0,$-m_0$)  \\[3pt]
      & (2/3,2/3,3/4) & (0,0,$m_0$) & (0,0,$m_0$)  \\[3pt]
\end{tabular}
\end{ruledtabular}
\end{table}

The new satellite peaks associated with the wavevector $\textbf{k}$ = ($\frac{1}{3}$, $\frac{1}{3}$, 0) that developed below 3~K have been attributed to the long-range magnetic ordering of the manganese atoms (Mn2) in the triangular layer. The neutron scattering data is well described by a magnetic structure with the $static$ magnetic moments pointing along the $c$-axis, and amplitudes which follow the $\textbf{k}$-wavevector modulation such as: $m_i=m_0 \Re[cos(2\pi\textbf{k}\cdot \textbf{r}_i+\phi)]$.\cite{magbook} Here, $m_0$ represents the amplitude of the ordered moment in the zero$th$ cell, and $\phi$ is a phase factor. The moment distribution for a phase $\phi$ = 0 along the $a$ axis is $m_0$, $-m_0$/2, $-m_0$/2, while for a choice of $\phi$ = $\pi$/2 the sequence becomes 0, $-\sqrt{3}/2 m_0$, $\sqrt{3}/2 m_0$. The moments are thus fully compensated inside the plane. Furthermore, the selection rule $L$=2$n$+1 indicates that successive triangular layers are stacked antiferromagnetically along the $c$ direction. It is important to realize that we are discussing the ordering of the $static$ moment. Other possible disordered or strongly-fluctuating magnetic components that may exist do not contribute to the Bragg intensities. The only possibility to obtain a uniform amplitude across all Mn2 magnetic sites is if an additional $\textbf{k}$ = (0,~0,~0) component would be present. This is important to understand, since most publications refer to this $intermediate$ state, which is specific to the triangular lattice systems with weak easy-axis anisotropy, as the $uud$ phase, without any further description of its single-$\textbf{k}$ or double-$\textbf{k}$ character. In our case, the lack of $\textbf{k}$ = (0,~0,~0) contribution (and thus of net magnetization) is obvious from the magnetization measurements. Furthermore, a careful measurement of the temperature evolution of the (1,0,1) reflection across the two low-temperature transitions, shown in Fig.~\ref{orderparam}, indicates no additional magnetic scattering as one would expect from an uncompensated magnetic component. Our amplitude modulated $uud$ (single-\textbf{k}) magnetic structure can be described by the magnetic space group $P6_3^\prime/m$ in a 3 x 3 x 1 magnetic supercell. Within this expanded unit cell, there are three non-equivalent magnetic sites (Mn2$_i$, $i$=1,2,3) that follow the modulation imposed by the $\textbf{k}$ wavevector. Note that the same magnetic space group can be used to describe a uniform $uud$ double-\textbf{k} structure. The proposed magnetic structure model for Mn2 site in the temperature range 2.2 K $\leqslant$ T $\leqslant$ 3 K is shown in Fig.~\ref{magstruct}(a), and the site-specific orientation of magnetic moments is given in Table~\ref{table3}. In Fig.~\ref{magstruct} we omitted the Mn1 honeycomb layers that preserve the same spin arrangement as shown in Fig.~\ref{honeymagstr}.

As evidenced by the order parameter curve in Fig.~\ref{orderparam}, an in-plane spin component develops below 2.2~K resulting in a canting of the ordered moment away from the $c$ direction. The in-plane spin component exhibits a modulation in amplitude in accord to $\textbf{k}$ = ($\frac{1}{3}$, $\frac{1}{3}$, 0). The natural tendency toward an uniform static moment magnitude for all sites is fulfilled by the development of in-plane components only on the two sites with reduced $m_c$ static moments. This can be realized by considering an offset between phase factors $\phi$ of the two moment components, $m_c$ and $m_{ab}$ of $\pi$/2. Ideally, this produces a 120$^\circ$ spin configuration in the plane containing the $c$-axis, and is known in literature as the planar ``Y'' canted structure. Similar to the intermediate phase, the successive Mn2 triangular layers remain aligned antiparallel with respect to each other. This low-temperature ordered state inherits the broken lattice symmetry while also breaking the spin-rotational symmetry within the $ab$ plane and thus can be viewed as an magnetic analogue to a supersolid state.\cite{seabra1,supersolid} The magnetic structure can be described using the monoclinic magnetic subgroup $P2_1^\prime$ (\#4.9) operating on a 3 x 3 x 1 magnetic supercell. We note that the in-plane spin component can align along any direction and, experimentally, this cannot be uniquely determined. For fitting the single-crystal neutron data we selected a model were the spins form an ideal 120$^\circ$ structure in the (1,1,0) plane. The refined magnitude of the static moment is 3.7(2)$\mu_B$, significantly smaller than that expected for spin $S$ = 5/2. Similar reduced ordered moment of about 3.9(5)$\mu_B$ was reported for the $S$=5/2 TLA system RbFe(MoO$_4$)$_2$.\cite{RbFeMoO_1} One possibility is that the static moment does not reach full saturation at the measured temperature of 1.7 K, but it could also be that the geometrical frustration are causing strong quantum fluctuations. The magnetic moments arrangement in this canted structural model for the temperatures below 2.2 K is summarized in Table~\ref{table3} and illustrated in Fig.~\ref{magstruct}(b).

\begin{figure*}[tbp]
\centering
\includegraphics[width=7in]{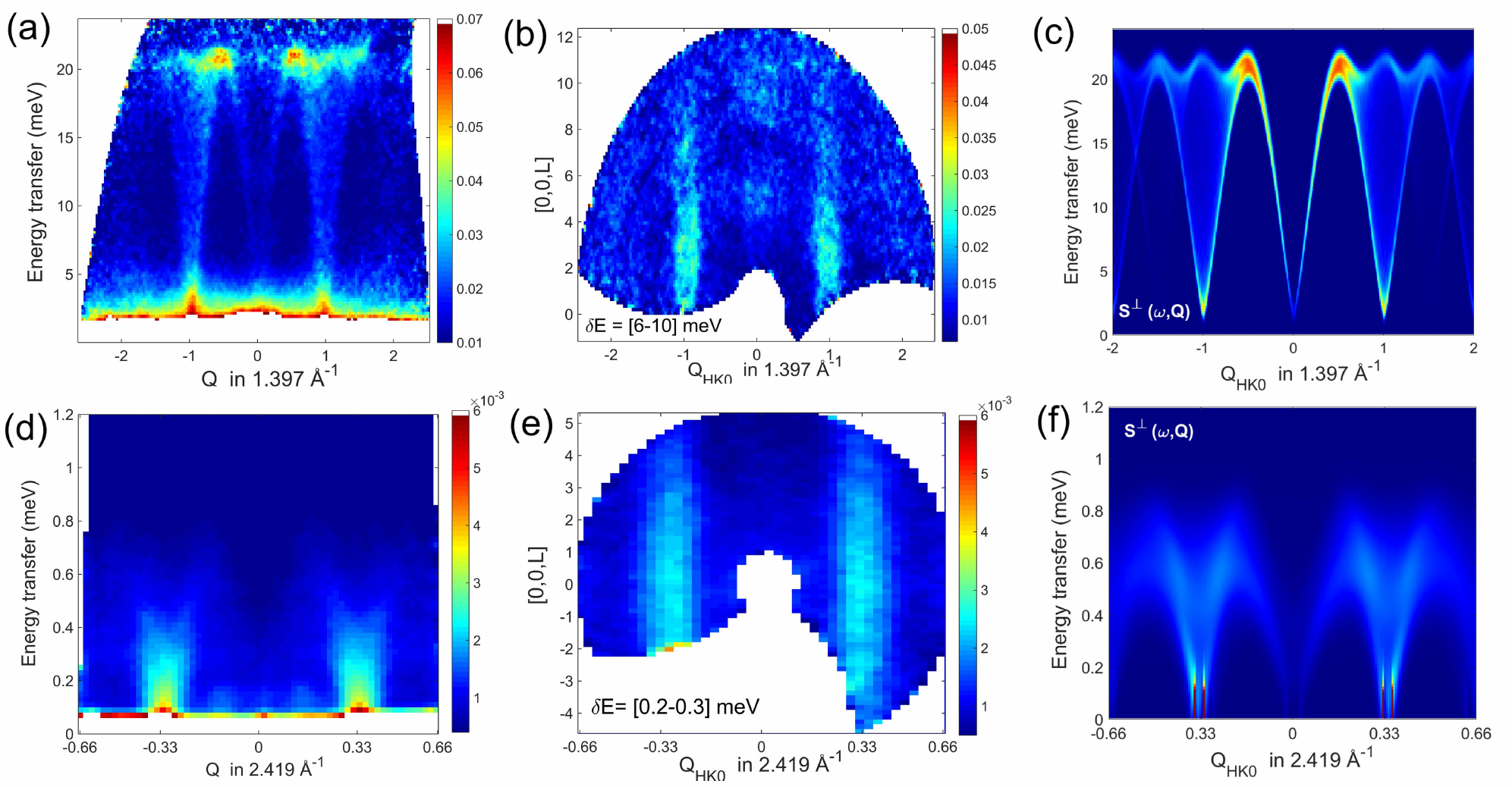}
\caption{\label{ins} (color online) (a) Energy-momentum slice of the inelastic neutron scattering data measured with $E_i$ = 25~meV at the HYSPEC spectrometer. A spin-wave branch emerges from the (1,0,0) magnetic peak and extends to an energy transfer of approximately 22 meV. (b) Contour plot of the ($H$,0,$L$) reciprocal plane for energy transfer integrated between 6 and 10 meV. The rod of scattering along [0,0,$L$] direction indicates very weak coupling between magnetic layers. (c) Calculated spin-wave spectrum using a Heisenberg model that includes first- and second-neighbor in-plane exchange interactions ($J_{1}$ = 1.55 meV, $J_{2}$ = -0.45(3) meV) in the S=5/2 honeycomb lattice. (d) Low-energy magnetic excitations that develop around the magnetic peaks at (1/3, 1/3, 1) and (2/3, 2/3,1). (e)  2D slice of momentum space corresponding to energy transfer range 0.2 - 0.3 meV revealing the quasi-two-dimensional character of the excitations. (f) The calculated spin-wave spectrum of S=5/2 triangular lattice Heisenberg antiferromagnet characterized by nearest-neighbor exchange interactions ($J$= 0.08 meV) and a easy-axis anisotropy ($D_t$ = 0.03meV). The spectrum is averaged over all in-plane \textbf{Q} directions to reproduce the experimental conditions.}
\end{figure*}

\subsection{Spin-wave excitations}

\subsubsection{Magnetic excitation spectrum of the honeycomb sublattice}

Inelastic neutron-scattering measurements performed with the incident energy E$_i$ = 25 meV revealed a well defined spin-wave branch emerging from the (1,0,0) magnetic peak and extending to an energy transfer of approximately 22 meV (see Fig.~\ref{ins}(a)). There is no discernable dispersion along the $c$-axis indicating very weak coupling between magnetic layers. As visible in Fig.~\ref{ins}(b), the two-dimensional correlations is evidenced by rods of scattering along [0,0,$L$] direction in the contour plot of the reciprocal lattice plane corresponding to the energy transfer range 6 - 10 meV. The scattering intensity along $\mathrm{\textbf{Q}}$ follows the decay expected for the magnetic form factor of Mn$^{2+}$ magnetic ion. As described in Section~\ref{Experiment} the sample used for these measurements was relatively small (\~ 0.3 g) and consisted of coaligned crystals only along $c$-axis direction with random in-plane orientation. This has negatively impacted the data quality and the amount of information that can be extracted from it. Thus, certain assumptions which simplified the analysis had to be made. Excitations from a honeycomb lattice are typically described by using a Heisenberg model that includes first-, second-, and third-neighbor in-plane exchange interactions ($J_{1}$, $J_{2}$, $J_{3}$), an exchange interaction between planes $J_{c}$, and an anisotropy term $D_h$.~\cite{honeycomb1,honeycomb2,swhoney}. Based on our experimental resolution we can estimate that the interplane coupling $J_{c}$ and the anisotropy are two order of magnitude smaller than the in-plane exchange interactions. The $J_{3}$ is not expected to have a significant contribution, and is beyond our ability to determine with the available data. The strength of nearest-neighbor exchange interaction can be estimated from the Curie-Weiss temperature ($\Theta_{CW}$ = -215 K), $J_{1}$ = 3$k_\mathrm{B}$ $\Theta_{CW}$/$\zeta$ $S$($S$+1).~\cite{Kittel} Considering the number of nearest-neighbor $\zeta$ = 3 and the spin value the $S$ =5/2 , one obtains $J_{1}$ = 2.1(3) meV. Fixing the $J_{1}$ value, the measured inelastic spectrum can be well reproduced by using an additional second-neighbor exchange $J_{2}$ = -0.3(1) meV. The ratio between the $J_{1}$ and $J_{2}$ is consistent with that expected for a N\'{e}el-type magnetic ground state in a honeycomb lattice.\cite{honeycomb1,honeycomb2}

In the absence of measurable spin-wave excitation gap, the magnitude of the axial anisotropy term $D_h$, responsible for the spins alignment along $c$-axis, can be obtained from the value of the field-induced transition observed near 11 T. For a honeycomb lattice ordered in a N\'{e}el-type AFM magnetic structure a spin-flip transition is expected to occur at a field $H_{sf}$ = 2S$\sqrt{D_h(3J_1-D_h)}$/g$\mu_B$. This yields an anisotropy $D_h$ of approximately 0.010(2) meV. The calculated spin-wave excitation spectrum using the aforementioned exchange parameters over an averaged in-plane momentum transfer is shown in Fig.~\ref{ins}(c). Our estimates show that the dipole-dipole interaction can present an important contribution to the uniaxial anisotropy of Mn$^{2+}$ ($S$ = 5/2) ions in the honeycomb layer. The intralayer and interlayer distances, 3.01~\AA~ and 5.5~\AA, yield an anisotropy comparable in magnitude (10$^{-2}$ meV) to those observed and calculated for the classical antiferromagnets MnO~\cite{Keffer1,Pepy} and MnF$_2$~\cite{Keffer2,Okazaki}, where the nearest-neighbors distances are about 3.1~\AA~ and 3.3~\AA, respectively.

\subsubsection{Magnetic excitation spectrum of the triangular sublattice}

The T = 1.7 K inelastic data contains additional low-energy magnetic excitations that are due to cooperative fluctuations of magnetically ordered spins in the triangular layer. As shown in Fig.~\ref{ins}(d) spin-wave excitations develop around the magnetic Bragg peak positions ($\frac{1}{3}$, $\frac{1}{3}$, 1) and ($\frac{2}{3}$, $\frac{2}{3}$, 1) and have a bandwidth of approximately 0.8 meV. The quasi-two-dimensional character of the excitations is revealed by the rod-like scattering along [0,0,$L$] direction in the 2D slice plot of momentum space corresponding to energy transfer range 0.2 - 0.3 meV, displayed in Fig.~\ref{ins}(e). Keeping in mind that only one-third of magnetic ions occupy the triangular layer, it is not surprising that the INS data quality for these low-energy excitation is even more affected by the small sample mass and its random in-plane orientation. In addition, a diffusive quasielastic component seems to overlap with the spin-waves, accounting for some of the missing ordered moment. Such diffuse scattering was previously observed in other triangular systems and was attributed to uncorrelated trimers that could survive well below the ordering temperature.~\cite{diffuse} Nonetheless, from the bandwidth of the spin-wave spectrum one can estimate the value of a nearest-neighbor exchange interaction ($J$) as approximately 0.08(1) meV. As pointed out in the previous sections, the magnetic behaviour observed from the triangular layer suggests the existence of a weak easy-axis anisotropy ($D_t$). The energy gap expected from such anisotropy  could not be resolved in our INS data. However, one can determine the value of $D_t$ from its relationship to the two-step transition temperatures $D_t \approx J$ ($T_{N2}$ - $T_{N3}$)/$T_{N2}$, or from the the saturation magnetic field of the triangular magnetic sublattice $H_{S}$ = (9$J$-2$D_t$)$S$/g$\mu_B$ = 14.3 T.~\cite{saturation} The estimated value of the axial anisotropy is $D_t$ $\approx$0.028(2) meV, which is nearly twice as large than that in the honeycomb layer $D_h$ $\approx$0.010(2) meV. Considering that the interatomic distance for Mn ions located in the triangular layer are much larger, $\sim$ 5.2 \AA, a much smaller dipolar contribution of the order of 10$^{-3}$ meV is estimated for this site. It therefore appears that the $D_t$ magnetic anisotropy is mainly due to higher-order terms in the interplay between crystalline-field and spin- orbit couplings and covalency effects, as demonstrated for other 3$d^5$ systems ~\cite{Francisco,Febbraro,Wan-Lun} For instance, spin-orbit coupling effects were shown to produce an anisotropy $D$ of about 0.06 meV in Mn$^{2+}$WO$_4$ .~\cite{Hollmann} One shall also stress that the trigonal bipyramidal coordination of Mn$^{2+}$ in the triangular layer is anticipated to induce a stronger spin-orbit-induced anisotropy than the more-regular octahedral environment of the magnetic ions in the honeycomb layer.~\cite{bipyramidal} The calculated in-plane averaged spin-wave spectrum using the estimated $J$ and $D_t$ parameters is shown in Fig.~\ref{ins}(f). The inelastic magnetic spectrum of the intermediate ordered state $uud$ has also been measured at 2.8~K and it shows no discernible change with respect to that observed at the base temperature.

\begin{figure*}[tbp]
\includegraphics[width=7in]{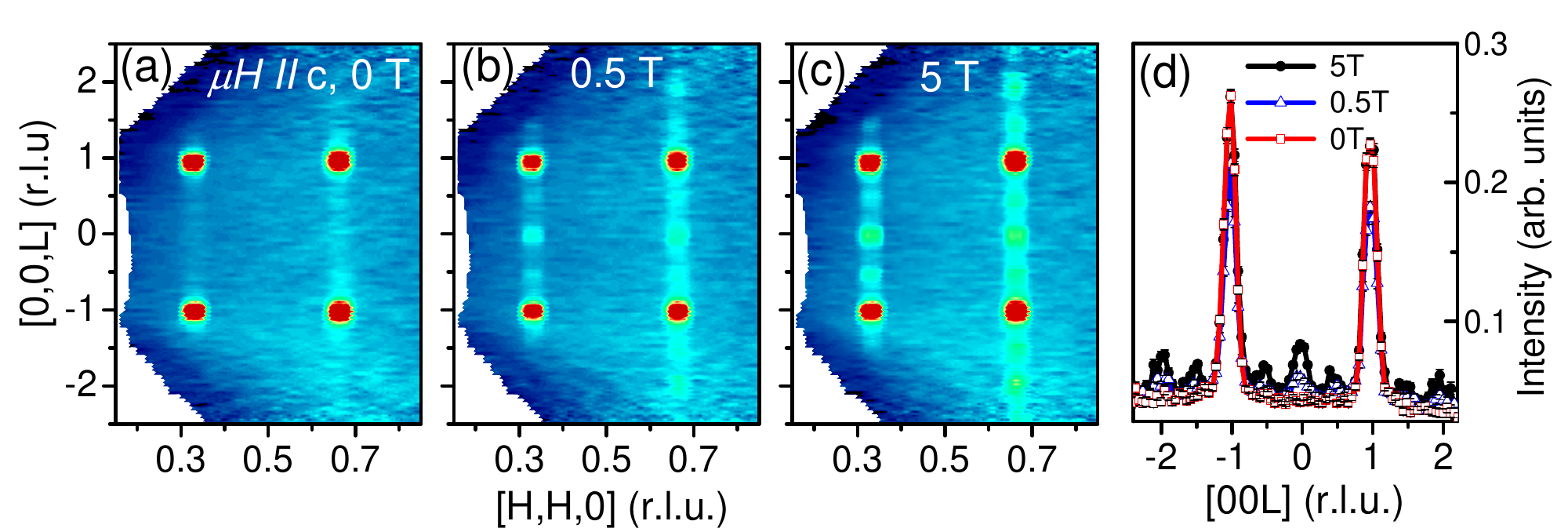}
\caption{\label{corelli_Hc} (Color online) (a-c) Evolution of scattering intensity in the ($H$,$H$,$L$) reciprocal plane for different magnetic fields applied along the $c$-direction. Data demonstrates the development of strong structured diffuse scattering intensity along the $L$-direction. (d) Cut along along the diffuse scattering at ($\frac{2}{3}$, $\frac{2}{3}$, $L$) position revealing prominent peaks centered at $L$ = 0, $\pm$0.5, $\pm$1 and $\pm$1.5.}
\end{figure*}

\begin{figure}[tbp]
\includegraphics[width=3.5in]{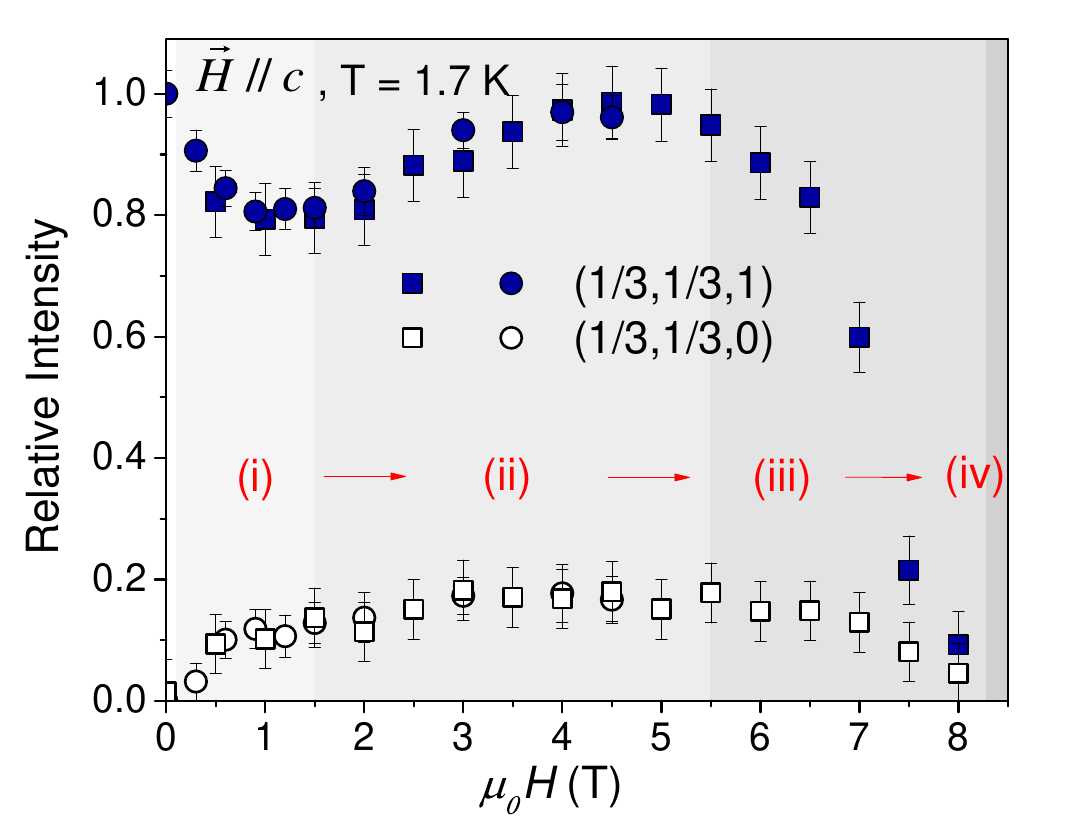}
\caption{\label{orderparam_Hc} (Color online) Field dependence of ($\frac{1}{3}$, $\frac{1}{3}$, 0) and ($\frac{1}{3}$, $\frac{1}{3}$, 1) peak intensities at T = 1.7 K and $\protect\overrightarrow{H}\parallel c$ exposing the presence of at least four intermediate magnetic states between the zero field ground state and the full magnetization saturation.}
\end{figure}

\begin{figure*}[tbp]
\includegraphics[width=7in]{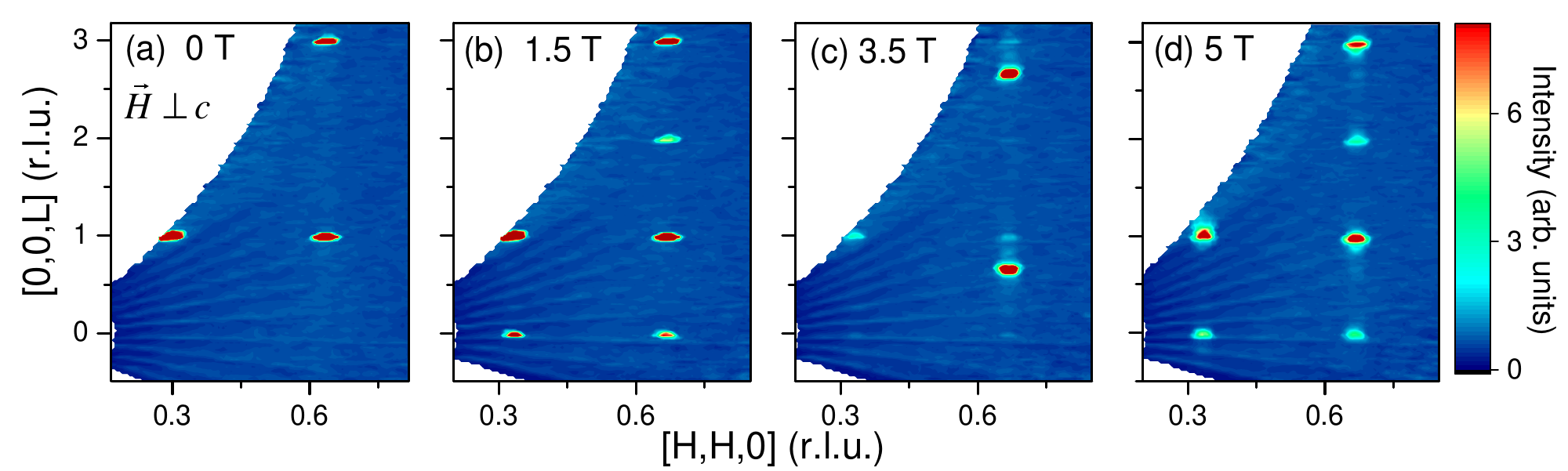}
\caption{\label{corelli_Hab} (Color online) (a-d) Representative plots of the ($H$,$H$,$L$) reciprocal plane obtained at T = 1.7 K for magnetic fields applied along the [1,$\overline{1}$,0] direction. The $\mu_0 H$ = 3.5 T data shows a shift along the $L$-direction of magnetic peaks to a slightly incommensurate position $\delta L$=0.654. At 5 T, the magnetic phase recovers its original commensurate state with the wavevector $\textbf{k}$ = ($\frac{1}{3}$, $\frac{1}{3}$, 0).}
\end{figure*}

\begin{figure}[tbp]
\includegraphics[width=3.5in]{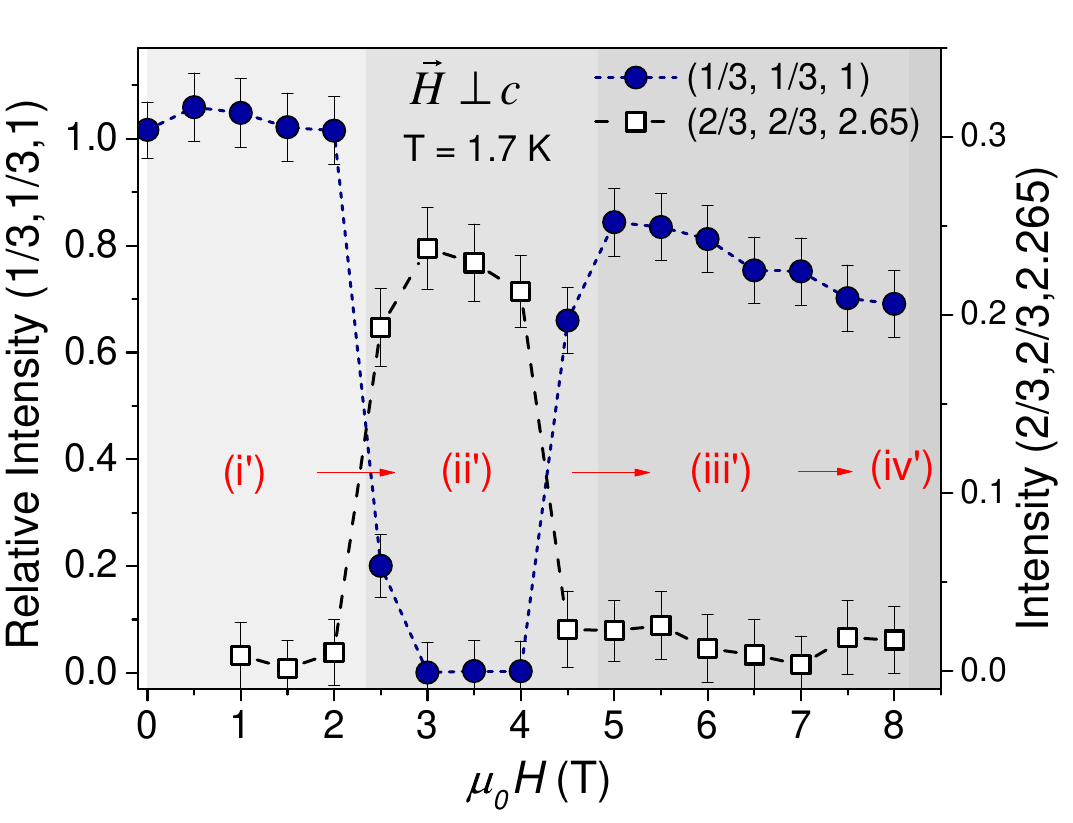}
\caption{\label{orderparam_Hab} (Color online) Field dependence of ($\frac{1}{3}$, $\frac{1}{3}$, 1) and ($\frac{2}{3}$, $\frac{2}{3}$, 2.654) magnetic peak intensities at T = 1.7 K mapping out the  phase diagram for $\protect\overrightarrow{H}\parallel$[1,$\overline{1}$,0]. One observes that the intermediate incommensurate phase separating two other field-induced commensurate states begins at about 2 T and disappears near 4.5 The dashed line is a guide to the eye.}
\end{figure}

\begin{figure}[tbp]
\includegraphics[width=3.4in]{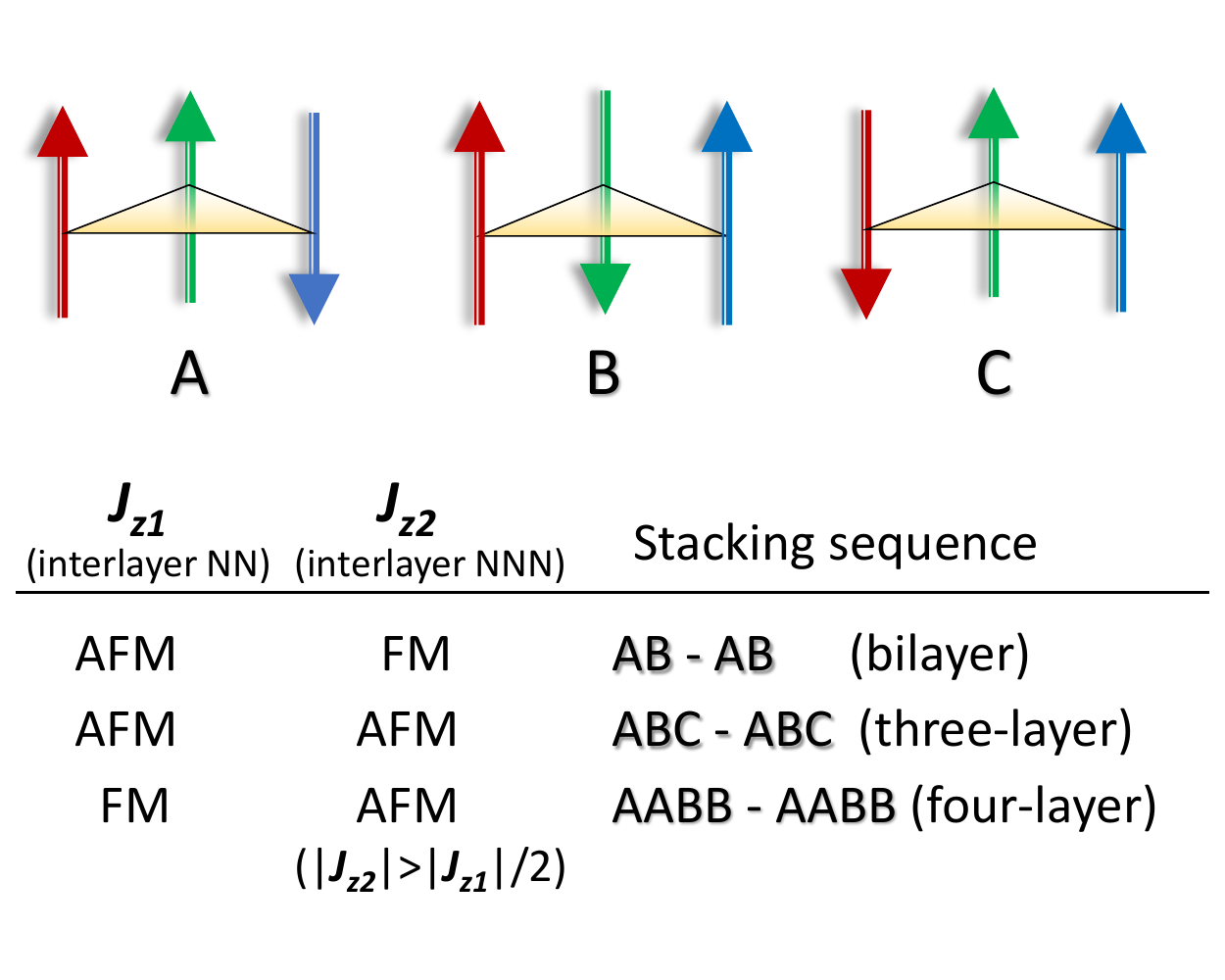}
\caption{\label{ABC_stacking} The three possible plateau state configurations due to geometric degeneracy of triangular layer: $A$ ($uud$), $B$ ($udu$) or $C$ ($duu$). The spin-stacking sequence of successive triangular layer is determined by the nature (antiferromagnetic (AFM) or ferromagnetic (FM)) and relative strengths of nearest-neighbor ($J_{z1}$) and second-nearest-neighbor ($J_{z2}$) interlayer interactions.}
\end{figure}

\subsection{Field-induced magnetic phases in triangular sublattice}

The effect of applied magnetic fields on the magnetic order of K$_{2}$Mn$_{3}$(VO$_{4}$)$_{2}$CO$_{3}$ has been studied using the CORELLI and HYSPEC instruments for fields up to 8 T. For such magnetic fields, only the triangular magnetic layer shows a notable magnetic response. Figure~\ref{corelli_Hc} shows the evolution of the elastic scattering in the ($H$,$H$,$L$) reciprocal plane for the magnetic fields of 0, 0.5 T and 5 T, applied along the $c$-axis direction. The data reveal changes in intensity of satellite magnetic peaks associated with the propagation vector $\textbf{k}$=($\frac{1}{3}$, $\frac{1}{3}$, 0), as well as the development of strong diffuse scattering intensity along the $L$-direction. The diffuse scattering becomes progressively stronger with increasing field, and has a well-structured profile. A cut along the $L$ direction of the diffuse scattering with prominent peaks at $L$=0, $\pm$0.5, $\pm$1 and $\pm$1.5 is displayed in Fig.~\ref{corelli_Hc}(d). A more thorough survey of the field effect on the intensities of ($\frac{1}{3}$, $\frac{1}{3}$, 0) and ($\frac{1}{3}$, $\frac{1}{3}$, 1) magnetic Bragg peaks has been carried out using HYSPEC instrument and is shown in Fig.~\ref{orderparam_Hc}. One can observe an immediate increase in intensity of ($\frac{1}{3}$, $\frac{1}{3}$, 0) peak as the field is raised above zero. Furthermore, the ($\frac{1}{3}$, $\frac{1}{3}$, 1) peaks intensity undergoes a slight decrease with a minimum near 1.5 T, followed by a recover with a local maximum at about 5.5 T, and then a strong decrease to a nearly complete disappearance at the highest measured field $\mu_0 H$ = 8 T. The profile of the field dependence reveals the existence of at least four different spin configurations before reaching the saturation.

When the magnetic field is applied along the [1,$\overline{1}$,0] direction, the evolution of magnetic peaks is strikingly different. Representative slices of the ($H$,$H$,$L$) reciprocal plane obtained at fields of 0, 1.5 T, 3.5 T and 5 T are presented in Figure~\ref{corelli_Hab}. The 1.5 T data reveals the appearance of well-defined ($\frac{1}{3}$, $\frac{1}{3}$, 0) and ($\frac{2}{3}$, $\frac{2}{3}$, 0) magnetic peaks and no diffuse scattering along the $L$-direction. At 3.5 T the magnetic peaks appear to relocate along $L$-direction to a slightly incommensurate lattice vector $\xi$ = 0.654(1). As the magnetic field is ramped up to 5 T, the magnetic phase recovers its commensurate state with $\textbf{k}$ = ($\frac{1}{3}$, $\frac{1}{3}$, 0). The order-parameter profiles of the commensurate and incommensurate magnetic states  as a function of the field at 1.7 K, are shown in Fig.~\ref{orderparam_Hab}. One can see that the incommensurate phase forms at about 2 T and disappears near 4.5 T. There is no detectable change in the incommensurability for this field range. Above 4.5 T, the commensurate ($\frac{1}{3}$, $\frac{1}{3}$, 1) peak reappears but does not recover all its intensity, and exhibits a slow decrease with increasing the field. The phase-diagram revealed by neutron scattering measurement sheds light on the origin of the split into two components of the heat capacity low-temperature peak, presented in Fig.~\ref{heatcapfield}. It is plausible that the heat capacity peak at $T_{N3}^*$ is related to the commensurate-incommensurate phase transition. The intermediate-field incommensurate state appears to be stable only over a finite temperature range, which suggests that  thermal fluctuations play a key role in its formation.

\begin{figure}[tbp]
\includegraphics[width=3.5in]{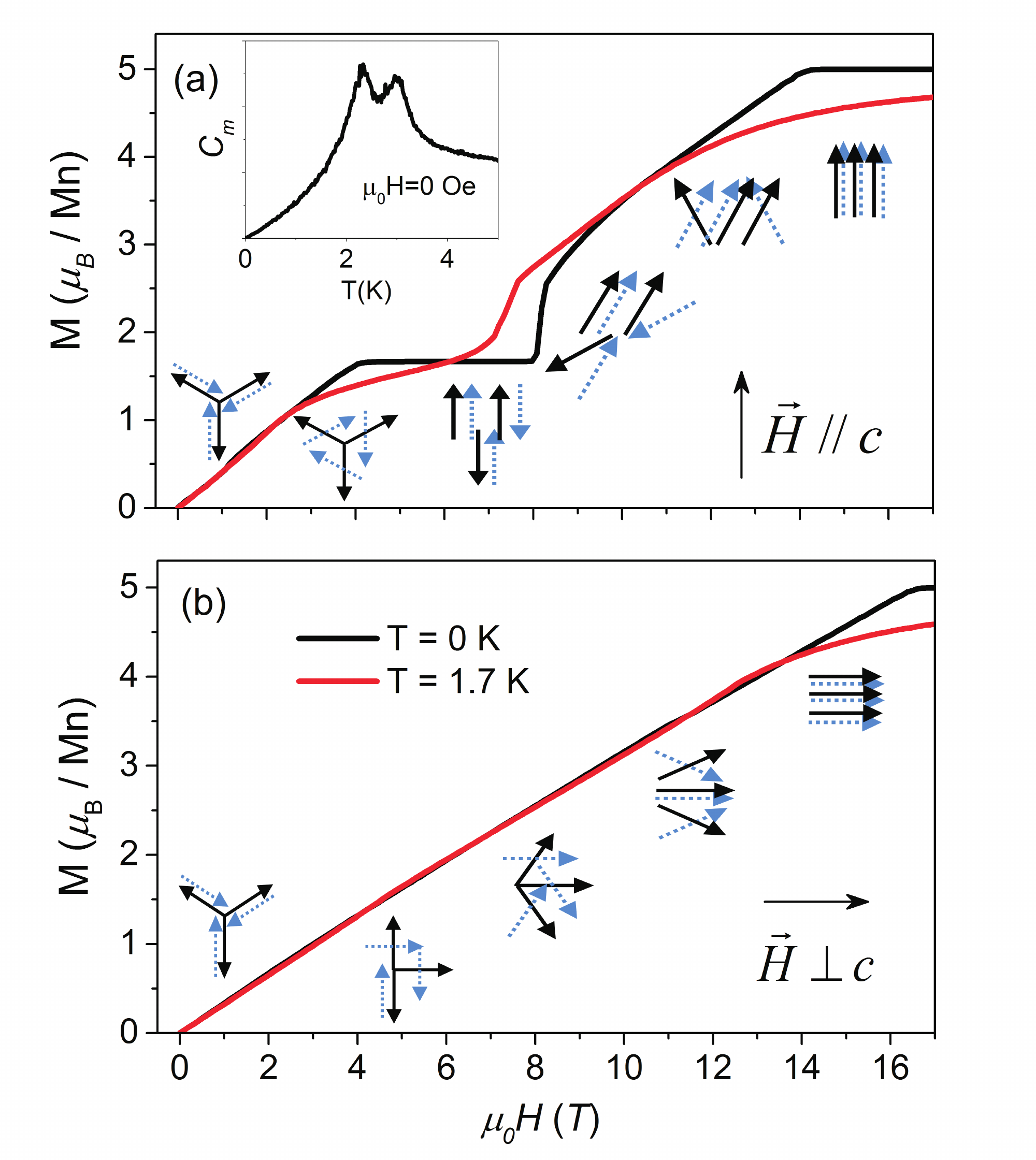}
\caption{\label{montecarlo} (Color online) Calculated magnetization curves for $\protect\overrightarrow{H}\parallel c$ (a) and $\protect\overrightarrow{H}\perp c$ (b) from a triangular magnetic lattice characterized by the NN exchange interaction and easy-axis anisotropy determined experimentally. Temperature effect on the magnetization plateau is shown by comparing the T = 0 K (black curve) with T = 1.7 K (red curve) calculations. The inset shows the computed zero-field heat-capacity curve that reproduces well the observed successive transitions 2.2 K and 3 K. Predicted field-induced bilayer spin states from our classical Monte Carlo simulations are sketched along the magnetization curves, with the planar $ac$ structures of the adjacent layers plotted on top of each other.}
\end{figure}

\begin{figure*}[btp]
\includegraphics[width=7in]{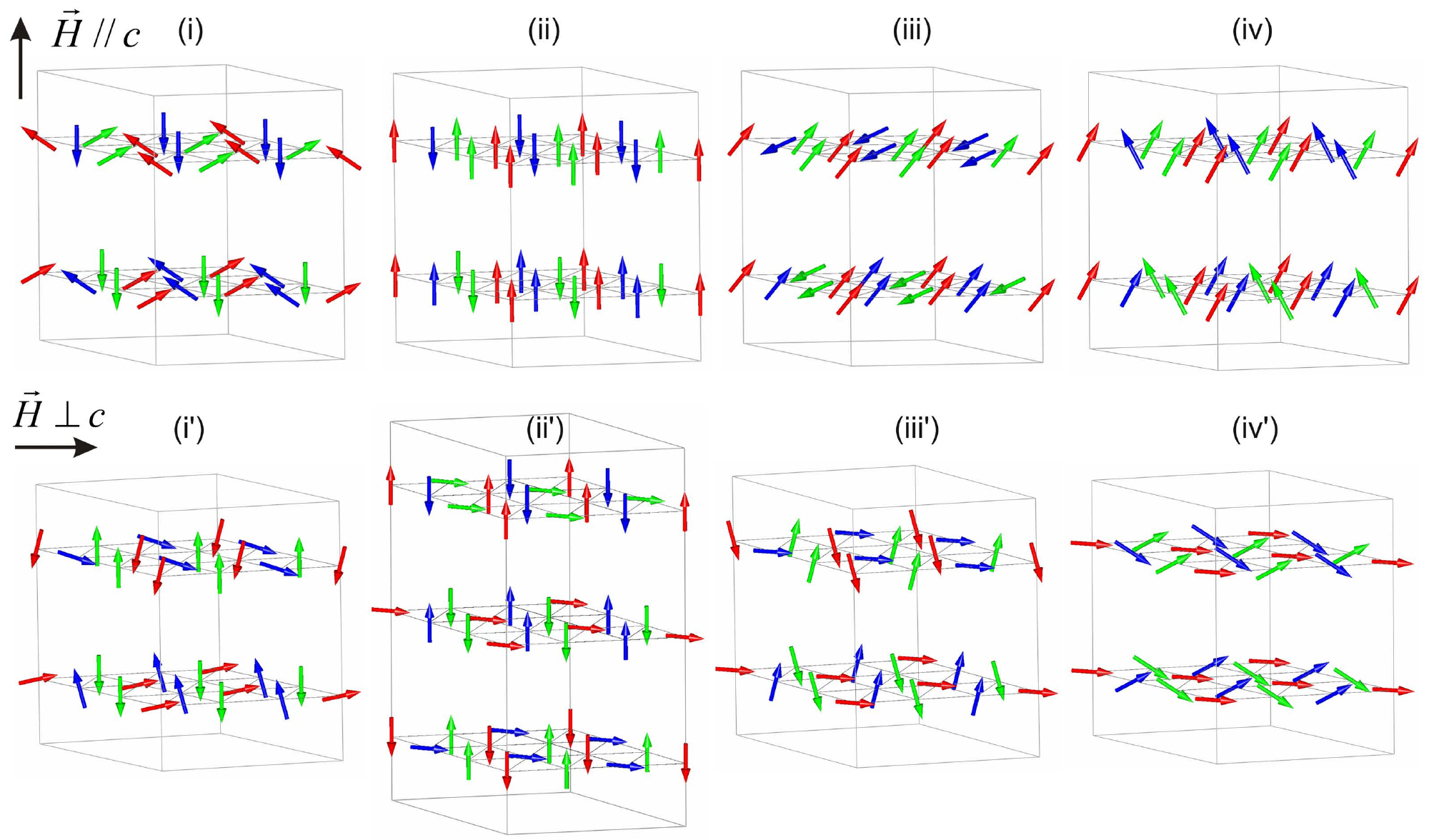}
\caption{\label{fieldmagstr} (Color online) Models of the static magnetic orders induced by magnetic fields applied parallel to $c$ (i-iv) and along [1,$\overline{1}$,0] (i'-iv') as determined from neutron diffraction data. The figure only presents the dominant bilayer structure but the (i-iii) states are susceptible to spin-stacking disorder and four-layer superstructure formation as described in the text. A three-layer stacking sequence (ii') was found for the in-plane field orientation where the moments form a nearly orthogonal configuration and are permuting orientations across the three layers.}
\end{figure*}

Based on the recorded integrated intensities, models of the magnetic structures corresponding to each stage of the $\overrightarrow{H}\parallel c$ or $\overrightarrow{H}\perp c$ phase-diagrams have been constructed. The available models were further scrutinized by considering the results of Monte-Carlo simulations performed on a spin Hamiltonian of a TLA described using the determined nearest-neighbor exchange coupling ($J$ = 0.08 meV) and easy-axis anisotropy ($D_t$ = 0.028 meV), as well as an approximate weak inter-layer coupling $J_{z1}$ = 0.005 meV. The calculated magnetization curves for T = 0 K and 1.7 K, along with the predicted magnetic configurations at 0 K are presented in Fig.~\ref{montecarlo}. The zero-field heat-capacity curve, shown in the inset of Fig.~\ref{montecarlo}, reproduces the two successive magnetic transitions at low-temperatures, while the calculated isothermal magnetization curve captures the experimental data very well.

The appearance of $L$ = even reflections in neutron data suggests the change in the spin-stacking sequence between triangular layers, with part of the spins tending to align  parallel to each other. Therefore, in the low-field regime of $\mu_0\overrightarrow{H}\parallel c$, the ``Y'' magnetic configurations of the adjacent triangular layers are no longer compensating each other as in the zero-field structure but follow the field direction to produce a net magnetization along the $c$-axis. The determined low-field magnetic structure is shown in Figs.~\ref{montecarlo}(a) and \ref{fieldmagstr}(i). It consists of two triangular layers, as the chemical crystal structure, with the stacked spins rotated by approximately 2$\pi$/3 relative one to another. Upon further increasing the magnetic field, the spins in each layer are continuously rotating to arrange parallel to the $c$-axis into a 1/3 $M_{ST}$ plateau state $uud$. This collinear configuration is displayed in Fig. \ref{fieldmagstr}(ii). Note that such ordered state has a double-$\textbf{k}$ nature, as $\textbf{k}$ = (0,~0,~0) contribution adds to the zero-field $\textbf{k}$ =($\frac{1}{3}$, $\frac{1}{3}$, 0). Due to the geometrical degeneracy, there are three possible magnetic configurations for each triangular layer: $uud$, $udu$ or $duu$, that we will refer to in the following as $A$ , $B$ and $C$ magnetic layers. These magnetic configurations are presented in Fig.~\ref{ABC_stacking}. The existence of stacking disorder of neighboring  layers is apparent in the presence of the structured diffuse scattering along the $L$-direction. The relative intensities at ($\frac{1}{3}$, $\frac{1}{3}$, $L$) peaks positions for $L$ = 0, 1 and 0.5 can be explained by an admixture of two magnetic polytypes: the bilayer $AB-AB$ and a four-layer superstructure $AABB-AABB$. The phase ratio of the two observed polytypes at T = 1.7 K is found to be nearly 4:1. Figure \ref{fieldmagstr} displays only the dominant $AB-AB$ bilayer magnetic polytype. While the spin-stacking disorder could be explained by the presence of small structural imperfections in the crystal, the four-layer polytype is indicative of the presence of second-nearest-neighbor inter-layer interactions. This is very surprising since the separation between second-nearest-neighbor triangular layers is very large ($\approx$~22.4~\AA), while each Mn atom is located in a zero-molecular-field of the adjacent honeycomb sublattice. As presented in Fig.~\ref{ABC_stacking}, a bilayer structure is expected for  AFM nearest-neighbor interlayer coupling ($J_{z1}$) and FM second-nearest-neighbor interlayer interaction ($J_{z2}$). In contrast, a four-layer superstructure $AABB-AABB$ is only possible for FM nearest-neighbor interlayer interaction ($J_{z1}$) weaker than twice the second-nearest-neighbor interlayer interaction ($2 \mid J_{z2} \mid > \mid J_{z1} \mid$). The experimental restraints prevented from carrying out measurement below 1.7 K to determine if the disorder and polytypes mixing ratio exhibits any temperature dependence. This might be expected if thermal and quantum fluctuations are involved in the effective interlayer coupling via order from disorder mechanism (effective interaction mediated by fluctuations of the magnetic moments in the honeycomb sublattice). The potential of quantum fluctuations to generate interlayer coupling via the mechanism of order from disorder has so far been considered for the case of the body-centered tetragonal lattice~\cite{Yildirim,orderbydisorder} but our experimental observations raise the possibility that similar mechanism can occur in triangular layers mediated by a honeycomb lattice. We recall that in our system the Mn ions at the triangular layer are located exactly on top or underneath the hollow center of the Mn honeycomb lattice. It is worth noting that evidence of two non-equivalent modulations with $L$ = 0 and $0.5$ and underlaying disorder has been also reported in the staggered three-layer TLA compound LiCrO$_2$.~\cite{LiCrO2}

At magnetic fields above 5.5 T the collinear state starts to evolve towards an oblique ``2:1'' structure, as shown in Fig.~\ref{fieldmagstr}(iii). As the field increases the magnetic moment that is pointing opposite to the field direction is continuously rotating to align with the field. In the absence of neutron data for fields larger than 8 T, we used classical Monte Carlo simulations to define the magnetic state. The calculations indicate that the moments are pass through a canted ``V'' spin configuration before the full saturation (M$_S$), as shown in Fig.~\ref{fieldmagstr}(iv). The field-induced magnetic configurations within the layer are consistent with the theoretical predictions and with those observed in similar triangular systems. The dominant bilayer magnetic polytypes (i),(ii) and (iii) found at low and intermediate fields are consistent with the configurations predicted by Gekht and Bondarenko for weakly-coupled triangular antiferromagnets.~\cite{gekht} However, the presence of multiple magnetic polytypes including the four-layer superstructure is indicative of more complicated interlayer coupling and calls for more thorough theoretical investigations.

In the case of the field applied perpendicular to $c$-axis, along the [1,$\overline{1}$,0] direction, the low-field magnetic structure can be looked as a distortion of the zero-field ``Y'' structure where the two canted spins are rotating towards the field direction. This reorientation does not alter the lattice periodicity along the $c$-direction and the magnetic structure remains bilayer up to about 2 T. The low-field magnetic structure that describes well the low-field (0 $< \mu_0H \leq$ 2 T) neutron data is shown in Fig.~\ref{fieldmagstr}(i$^\prime$). The rotation of the spins continues until they reach a nearly-orthogonal arrangement, at about 2 T, with one spin pointing along the in-plane field direction and the other two aligned nearly parallel to $c$-axis (i.e. \textit{up-right-down} or $urd$ magnetic structure). Similarly to the plateau case described by $A$, $B$, $C$ configurations (Fig.~\ref{ABC_stacking}), any permutation of the three spin orientations is possible for such an orthogonal model. Our diffraction data clearly indicates that a three-layer superstructure is stabilized at intermediate fields between approximately 2 T and 4.5 T. This superstructure is shown in Fig.~\ref{fieldmagstr}(ii$^\prime$). The stacked spins are alternating their orientations across the three layers as, $udr$, $rud$ and $dru$, equivalent to a $ABC-ABC$ configuration that becomes stable when both nearest-neighbor and second-nearest-neighbor are antiferromagnetic. There is, however, a small incommensurability of the modulation along $c$-axis ($\textbf{k}$=($\frac{1}{3}$, $\frac{1}{3}$, 0.654)), and therefore the spin orientations undergo slight undulations around the ideal directions. This could arise from a competition between nearest-neighbor and second-nearest-neighbor inter-layer interactions. It is interesting to point out that the $c$-axis modulation of the magnetic order appears to be locked-in to a constant value for the entire range (2 $<\mu_0H <$ 4.5 T), despite the fact that the magnetization curve shows a uniform increase suggesting a continuum rotation of the spins towards the field direction. It is also important to note that no stacking disorder is observed, which indicates that this intermediate-field incommensurate magnetic phase is well stabilized. Our classical Monte Carlo simulation, which only accounts for a weak nearest-neighbor interlayer exchange, predicts for this intermediate-field range only a bilayer spin arrangement in the form of $udr$ -- $rud$ (see Fig.~\ref{montecarlo}(b)). Considering the stability of the three-layer phase over a finite temperature interval, one may infer that the second-nearest-neighbor interlayer coupling are strengthened by quantum-mechanical coupling via order-by-disorder mechanism.~\cite{henley,Yildirim,orderbydisorder}. Upon increasing the magnetic field above 4.5 T the magnetic structure recovers its original bilayer form, with a tilted variant of $udr$ -- $rud$ configuration, labeled as ``W'', illustrated in Fig.~\ref{fieldmagstr}(iii$^\prime$). Our Monte Carlo calculation predicts that before saturation the spins may experience an additional change in the stacking pattern as drawn in Figs.~\ref{montecarlo}(b)and ~\ref{fieldmagstr}(iv$^\prime$).

\section{Summary}

The structural and magnetic properties of the vanadate − carbonate K$_{2}$Mn$_{3}$(VO$_{4}$)$_{2}$CO$_{3}$ have been studied by means of
magnetization, specific-heat and neutron scattering measurements. The structure consists of an alternate stacking of honeycomb and triangular layers made of edge sharing MnO$_6$ octahedra and MnO$_5$ trigonal-bipyramids, respectively. Contrary to what was previously reported in Ref.~\onlinecite{Yakubo}, and in agreement with the first-principles calculations of Ref.~\onlinecite{Sama}, we found that both layers consists of Mn$^{2+}$ in in high-spin state. The two magnetic layers act as nearly independent magnetic sublattices with magnetic interactions of completely different energy scales. The honeycomb magnetic sublattice orders at about 85 K in a Ne\'{e}l-type AFM magnetic structure, with the Mn moments oriented parallel to the $c$-axis. This AFM state produces a zero-molecular-field on Mn atoms located in the triangular sublattice, which order magnetically in two steps at much lower temperatures. Analysis of neutron diffraction data show that the first transition, at $\approx$ 3 K, is to an amplitude-modulated AFM collinear $uud$ magnetic structure described by a propagation vector $\textbf{k}$ = ($\frac{1}{3}$, $\frac{1}{3}$, 0). The second magnetic transition, at 2.2 K, corresponds to the development of in-plane spin components that leads to the formation of a canted ``Y'' magnetic structure in a plane containing the $c$-axis. Successive triangular layers are aligned antiferromagnetically along the $c$-direction. Magnetization measurements performed under applied magnetic fields revealed a 1/3 magnetization plateau and a saturation near 14.3 T of the triangular sublattice, and a spin-flop reorientation of the honeycomb sublattice at about 11 T. On the basis of these field-induced transitions, the axial magnetic anisotropy of the octahedrally coordinated Mn ions in the honeycomb lattice was estimated to be 0.010(2) meV, while the anisotropy in the pentagonal-bipyramidal environment of the triangular lattice was found to be appreciably larger, about 0.028(2) meV. Inelastic scattering measurements revealed spin-wave excitations with a strong two-dimensional character. The excitation branch associated to the honeycomb ordered state was described by using a Heisenberg model that includes first-neighbor $J_{1}$ = 2.1(3) meV, and second-neighbor $J_{2}$ = -0.3(1) meV in-plane exchange interactions. On the other hand, the spin-wave spectrum of the triangular sublattice yielded an estimate of the nearest-neighbor exchange interaction value of 0.08(1) meV. The determined exchange interactions were used for Monte Carlo simulations to obtain the temperature -- magnetic field phase diagram. A systematic neutron diffraction study for applied magnetic-field along the $c$-axis revealed that the triangular magnetic lattice undergoes at least four intermediate magnetic phases before reaching the saturation. These phases are generally consistent with the previous theoretical predictions for TLA and with our Monte Carlo simulations and include the canted ``Y'', collinear $uud$, canted ``2:1'' and ``V'' in-layer spin configurations. Nevertheless, some discrepancies have been identified with regard to the spin-stacking sequence of neighboring layers. A stacking disorder and a mixture of bilayer and four-layer magnetic polytypes was revealed by the presence of highly structured magnetic diffuse scattering along the $L$-direction. An applied magnetic field perpendicular to $c$-axis is also found to produce at intermediate fields a novel magnetic state that exhibits a three-layer periodicity along the $c$-direction. In the three-layer structure the spins are alternating orientations in a quasi-orthogonal arrangement with two perpendicular and one parallel to the field direction. The appearance of such magnetic superstructures indicate that at finite temperature and intermediate fields the second nearest-neighbor interlayer interactions cannot be ignored, and may become instrumental in generating new magnetic orderings. This raises the possibility that subtle thermal and quantum fluctuations may generate effective interlayer tunneling between the triangular layers separated by honeycomb lattices. Furthermore, the formation of multiple spin-stacking sequences in a single material is very remarkable since each superstructure requires different types (AFM or FM) and different relative strengths of the interlayer nearest-neighbor and second-nearest-neighbor interactions. We hope that our findings will stimulate further theoretical studies and launch new considerations of field-induced magnetic phase diagrams in weakly-coupled triangular lattices.

\begin{acknowledgments}
Work at the Oak Ridge National Laboratory, was sponsored by US Department of Energy, Office of Science, Basic Energy Sciences, Scientific User Facilities Division (neutron scattering) and Materials Sciences and Engineering Division (magnetization and heat capacity analysis). The authors also acknowledge the financial support from the National Science Foundation under grant no. DMR-1410727. A portion of this work was performed at the National High Magnetic Field Laboratory (NHMFL), which is supported by NSF Cooperative Agreement No. DMR-1157490 and the State of Florida. We acknowledge Stan Tozer for use of his 16 T PPMS which is supported by the NHMFL and the Center for Actinide Science and Technology (CAST), an EFRC funded by the DOE - BES under Award Number DE-SC0016568. VOG thanks Xiaoping Wang for the help provided with the TOPAZ single crystal data refinement, and Igor Zaliznyak for useful discussions.

Notice: This manuscript has been authored by UT-Battelle, LLC under Contract No. DE-AC05-00OR22725 with the U.S. Department of Energy. The United States Government retains and the publisher, by accepting the article for publication, acknowledges that the United States Government retains a non-exclusive, paid-up, irrevocable, world-wide license to publish or reproduce the published form of this manuscript, or allow others to do so, for United States Government purposes. The Department of Energy will provide public access to these results of federally sponsored research in accordance with the DOE Public Access Plan (http://energy.gov/downloads/doe-public-access-plan).
\end{acknowledgments}

\end{document}